\documentclass[aps,prfluids, superscriptaddress,nofootinbib]{revtex4-2}

\usepackage[a4paper,margin=1in]{geometry}
\usepackage{amsmath,amssymb,bm,mathtools,amsthm}
\usepackage{xcolor}
\usepackage{hyperref}
\usepackage{enumitem}
\usepackage{graphicx}
\usepackage{booktabs}
\usepackage{microtype}
\usepackage{CJKutf8}

\hypersetup{colorlinks=true,linkcolor=blue,citecolor=blue,urlcolor=blue}

\newcommand{\dd}{\mathrm{d}}
\newcommand{\D}{\mathrm{D}}
\newcommand{\grad}{\bm{\nabla}}
\newcommand{\Div}{\bm{\nabla}\!\cdot}
\newcommand{\Curl}{\bm{\nabla}\!\times}
\newcommand{\eps}{\varepsilon}
\newcommand{\bbI}{\mathbb{I}}
\newcommand{\Id}{\mathrm{Id}}
\newcommand{\SO}{\mathrm{SO}(3)}
\newcommand{\tr}{\mathrm{tr}}
\newcommand{\dev}{\mathrm{dev}}
\newcommand{\STF}[1]{\left\langle #1 \right\rangle}
\newcommand{\avg}[1]{\left\langle #1 \right\rangle}
\newcommand{\vort}{\bm{\zeta}}
\newcommand{\fext}{\bm{F}^{\mathrm{ext}}}
\newcommand{\tauext}{\bm{\tau}^{\mathrm{ext}}}
\newcommand{\eh}{\mathrm{h}}
\newcommand{\ax}{\mathrm{ax}}
\newcommand{\intc}{\mathrm{int}}
\newcommand{\mech}{\mathrm{mech}}

\definecolor{HLORANGE}{RGB}{213,94,0}
\definecolor{HLGREEN}{RGB}{0,125,80}
\makeatletter
\def\HLTXT{%
  \@ifnextchar_{\HLTXT@dispatch}{\HLTXT@error}}
\def\HLTXT@dispatch_{%
  \@ifnextchar B{\HLTXT@blue}{%
    \@ifnextchar O{\HLTXT@orange}{%
      \@ifnextchar G{\HLTXT@green}{\HLTXT@error}}}}
\def\HLTXT@blue BLUE#1{{\color{black}#1}}
\def\HLTXT@orange ORANGE#1{{\color{black}#1}}
\def\HLTXT@green GREEN#1{{\color{black}#1}}
\def\HLTXT@error#1{\PackageError{main}{Unsupported HLTXT_* macro usage}{Use \string\HLTXT_BLUE\{...\}, \string\HLTXT_ORANGE\{...\}, or \string\HLTXT_GREEN\{...\}.}}
\makeatother

\begin{document}

\title{Retained-spin micropolar hydrodynamics from the Boltzmann--Curtiss equation}
\author{Satori Tsuzuki (\begin{CJK}{UTF8}{ipxm}都築怜理\end{CJK})}
\affiliation{Research Center for Advanced Science and Technology, The University of Tokyo}

\begin{abstract}
We derive a retained-spin micropolar hydrodynamic closure from the Boltzmann--Curtiss equation using a generalized Chapman--Enskog construction in which the local mean spin is retained as a quasi-slow variable. Starting from the one-particle kinetic balance identities and the corresponding exact coarse-grained finite-size balances for mass, linear momentum, and intrinsic angular momentum, we keep the collisional-transfer contribution to the antisymmetric stress explicit in the spin balance, decompose the first-order source into irreducible scalar, axial, and symmetric-traceless sectors, and show explicitly how the standard micropolar constitutive structure with coefficients $(\eta,\xi,\eta_r,\alpha,\beta,\gamma)$ emerges. This decomposition makes clear that the one-particle kinetic stress contributes only to the symmetric stress, whereas the rotational viscosity belongs to a collisional-transfer channel. For perfectly rough elastic hard spheres, we further obtain explicit dilute-gas estimates for the rotational viscosity $\eta_r$ from homogeneous spin relaxation and for the transverse spin-diffusion combination $\beta+\gamma$ from a transport-relaxation calculation. Targeted event-driven molecular-dynamics simulations are used as a posteriori checks: expanded homogeneous-spin density and roughness sweeps support the predicted $n^2$ and $K/(K+1)$ trends for $\eta_r$, while finite-$k$ transverse runs provide a qualitative diagnostic of the retained-spin response. The result is a self-contained derivation and coefficient-level estimate of retained-spin micropolar hydrodynamics that clarifies which parts of the closure are exact balance-law statements, which are first-order generalized Chapman--Enskog results, and which remain controlled rough-sphere estimates.
\end{abstract}

\date{\today}

\maketitle


\section{Introduction}
Micropolar and spin-carrying continuum theories extend classical fluid
mechanics by promoting local micro-rotation to an independent field alongside
the mass density, velocity, and temperature
\cite{DahlerScriven1961,CondiffDahler1964,Eringen1966,Eringen1999,Eringen2001}.
At the continuum level this introduces an antisymmetric stress channel, a
couple stress, and a characteristic relaxation between vorticity and local
spin. At the kinetic level the same physics is traced back to binary collisions
that exchange translational and rotational angular momentum
\cite{CurtissDahler1963,DahlerSather1963,MonchickYunMason1963,CondiffLuDahler1965,McCoySandlerDahler1966}.

The basic links between the kinetic and continuum descriptions are classical,
but in practice the derivation is scattered across several literatures. The
exact moment balances are often stated without algebra. The Chapman--Enskog
construction is typically described only in outline. The antisymmetric stress
channel, which is the key to rotational viscosity, is especially susceptible to
confusion because it is not generated by the symmetric one-particle kinetic
stress and, for finite-size rough particles, must already be kept explicit in
the exact coarse-grained intrinsic-spin balance. In addition, when a local spin
field is retained at the hydrodynamic level, the calculation is no longer a
strict hydrodynamic reduction in the narrow sense: the mean spin is not a
collision invariant in general and must be treated as a quasi-slow variable in
the sense of extended thermodynamics
\cite{MullerRuggeri1998,JouCasasLebon1999}.

The present derivation should be viewed against a literature that is real but
highly dispersed. On the continuum side, early work on polar and spin-carrying
fluids identified the antisymmetric stress and angular-momentum balance
structure that later became standard in micropolar theory
\cite{DahlerScriven1961,CondiffDahler1964,Eringen1966}. On the kinetic side,
the Boltzmann--Curtiss and loaded-sphere programs introduced generalized
kinetic equations with rotational degrees of freedom
\cite{CurtissDahler1963,DahlerSather1963}, while rough-sphere transport and
Chapman--Enskog analyses were developed for dilute gases
\cite{MonchickYunMason1963,CondiffLuDahler1965} and then extended to denser or
more general polyatomic-fluid settings
\cite{McCoySandlerDahler1966,TheodosopuluDahler1974I,TheodosopuluDahler1974II,DahlerTheodosopulu1975}.
These works provide many of the formal ingredients used below, but the relevant
steps are scattered across continuum mechanics, molecular kinetic theory, and
dense-fluid or rough-sphere transport theory rather than presented as a single,
detailed retained-spin derivation.

\HLTXT_ORANGE{Earlier kinetic formulations for nonspherical molecules go back to Curtiss's classical treatment, which was later developed into the Boltzmann--Curtiss and loaded-sphere programs \cite{Curtiss1956}. On the continuum side, standard micropolar-fluid theory and applications are described not only in Eringen's monographs but also in the mathematical and applied treatment of Lukaszewicz \cite{Lukaszewicz1999}. In granular mechanics, Babic's averaged balance equations provide an important coarse-grained perspective on stress and couple stress \cite{Babic1997}, while boundary-driven and inclined-plane granular flows illustrate why micropolar effects are often most visible near walls or shear-localized regions \cite{Mitarai2002}.}

Subsequent developments also proceeded largely along separate tracks.
Micropolar continuum theory matured as an independent constitutive framework
\cite{Eringen1999,Eringen2001}; granular and rough-sphere kinetic theories
emphasized rapid granular flow, rough-particle transport, and hydrodynamic
closures for particles with rotational degrees of freedom
\cite{Lun1991,Hayakawa2003,MegiasSantos2021}; and recent
Boltzmann--Curtiss-based studies have focused either on local-spin first-order
closures or on higher-order constitutive models for polyatomic gases
\cite{WonnellChen2018,SinghKarchaniSharmaMyong2020,MankodiMyong2020}. To our
knowledge, however, a self-contained account that combines the exact balance-law structure,
a retained-spin Chapman--Enskog construction, an irreducible first-order sector
decomposition, and explicit dilute-gas rough-sphere estimates for the
rotational coefficients has not been readily available in one place. The role
of the present paper is therefore primarily reconstructive and expository: it
collects these strands into a single notation and makes explicit which parts of
the retained-spin closure are exact, which are formal Chapman--Enskog results,
and which remain estimate-level rough-sphere evaluations.

\HLTXT_ORANGE{The novelty claimed here is therefore not a new rough-sphere collision model and not a complete numerical validation of all micropolar transport coefficients. The contribution is the combination of four elements in a single notation. First, the one-particle moment identities are placed next to the exact finite-size balance laws, so that the antisymmetric contact-transfer torque is visible before constitutive assumptions are made. Second, the mean spin is retained through an explicitly stated extended-hydrodynamic ordering rather than silently treated as a collision invariant. Third, the first-order source is decomposed into irreducible scalar, axial, and symmetric-traceless sectors, making clear that the axial stress channel cannot come from the symmetric one-particle kinetic stress. Fourth, for a transparent rough-sphere reference model, low-density estimates are given for $\eta_r$ and for the transverse combination $\beta+\gamma$. The paper is thus primarily a structural derivation and coefficient-estimate paper.}

The aim of the present paper is therefore narrow but useful. We do not attempt
to develop a new response theory or a new application. The numerical material added here remains deliberately targeted rather
than exhaustive: expanded EDMD benchmarks are used only to test the dilute
rough-sphere estimates rather than to launch a separate transport-simulation
program. We otherwise write out a detailed derivation of
the retained-spin closure from the Boltzmann--Curtiss equation, with special
attention to the exact balance-law structure, the irreducible tensor decomposition of
the first-order source, the formal first-order coefficient problem, and the
dilute-gas estimates of the key rotational coefficients for perfectly rough
elastic hard spheres. The paper is intended to be readable line-by-line by a
reader who wants to check the intermediate manipulations.
A companion manuscript Ref.~\cite{Tsuzuki2026CompanionRespTheoConseq} addresses 
the response-theoretic consequences of the retained-spin closure, including EDMD 
observability and model discrimination among retained-spin, eliminated-spin, and 
polynomial higher-gradient descriptions. The present paper is instead 
derivation- and coefficient-estimate-oriented: its purpose is to provide 
the exact balance-law structure with the collisional-transfer torque channel 
made explicit, the generalized Chapman--Enskog construction with 
retained spin, and dilute-gas rough-sphere estimates for selected key rotational coefficients.

The scope is the following.
\begin{enumerate}[label=(\roman*)]
\item We derive the one-particle kinetic balance identities from the
Boltzmann--Curtiss equation and write the corresponding exact finite-size
intrinsic-spin balance with the collisional-transfer stress torque made
explicit.
\item We formulate a generalized Chapman--Enskog expansion retaining the local
mean spin $\bm{\omega}_0$ as a quasi-slow variable.
\item We decompose the first-order source into irreducible scalar, axial, and
symmetric-traceless sectors and write the corresponding formal integral
equations for the response functions.
\item We show explicitly how the standard micropolar constitutive equations are
recovered at first order, and we identify the precise place where the
rotational viscosity $\eta_r$ enters.
\item For perfectly rough hard spheres, we derive explicit dilute-gas estimates
of $\eta_r$ and of the transverse spin-diffusion combination $\beta+\gamma$.
\item We supplement these rough-sphere estimates with targeted EDMD checks of
homogeneous spin relaxation, including density and roughness sweeps, and of a
finite-$k$ transverse retained-spin mode.
\end{enumerate}

Two limitations should be stated at the outset. First, the present
Chapman--Enskog construction is \emph{generalized} or \emph{extended}: the mean
spin is treated as quasi-slow rather than as a strict collision invariant. More
precisely, the retained-spin ordering adopted below assumes that the residual
axial relaxation of the retained-spin manifold, represented in the
collisional-transfer channel, is $O(\eps)$, i.e. of the same asymptotic order
as the first gradient corrections. Second, the antisymmetric stress channel is
structurally closed at first order, but a fully coefficient-complete evaluation
of $\eta_r$ for a rough-sphere collision operator would require an explicit
collisional-transfer bracket. The rough-sphere formulas derived here should
therefore be read as controlled low-density estimates rather than as the final
word on the full antisymmetric Chapman--Enskog problem.

\section{Kinetic setting and macroscopic moments}
\subsection{State space and Boltzmann--Curtiss equation}
We consider a dilute gas of rigid particles undergoing binary collisions. A
single particle is characterized by its translational velocity
$\bm{v}\in\mathbb{R}^3$ and intrinsic angular velocity
$\bm{\omega}\in\mathbb{R}^3$. For perfectly rough spheres the orientation is
irrelevant and the one-particle distribution is
$f(\bm{x},\bm{v},\bm{\omega},t)$. For general rigid rotators an orientation
$R\in\SO$ must also be included and the distribution becomes
$f(\bm{x},\bm{v},R,\bm{\omega},t)$. We write the phase-space measure as
\begin{equation}
\dd\Gamma=
\dd^3v\,\dd^3\omega
\quad\text{(rough spheres)},
\qquad
\dd\Gamma=
\dd^3v\,\dd R\,\dd^3\omega
\quad\text{(general rigid rotators)}.
\end{equation}

A convenient working form of the Boltzmann--Curtiss equation is
\begin{equation}
\partial_t f
+v_j\partial_{x_j}f
+\frac{F^{\mathrm{ext}}_j}{m}\partial_{v_j}f
+(\bbI^{-1}\tauext)_j\partial_{\omega_j}f
+\mathcal{R}[f]
=C[f,f].
\label{eq:BC}
\end{equation}
Here $m$ is the particle mass, $\bbI$ is the inertia tensor, $\fext$ is the
external force on a particle, $\tauext$ is the external torque about the
particle center, $\mathcal{R}[f]$ is the orientation-streaming term relevant
for rigid rotators, and $C[f,f]$ is the binary collision operator. For
perfectly rough spheres, $\mathcal{R}[f]=0$. We assume rapid decay of $f$ in
$\bm{v}$ and $\bm{\omega}$ so that the integrations by parts below have no
boundary contributions. The integration over $R\in\SO$ contributes no boundary
terms because the Haar measure is invariant and the manifold is compact.

\subsection{Macroscopic fields and peculiar variables}
The number density, mass density, mean velocity, and mean spin are defined by
\begin{align}
n(\bm{x},t)&:=\int f\,\dd\Gamma,
&
\rho&:=mn,
\label{eq:def_nrho}\\
u_i(\bm{x},t)&:=\frac{1}{n}\int v_i f\,\dd\Gamma,
&
\omega_{0i}(\bm{x},t)&:=\frac{1}{n}\int \omega_i f\,\dd\Gamma.
\label{eq:def_uomega}
\end{align}
The associated peculiar variables are
\begin{equation}
c_i:=v_i-u_i,
\qquad
\Omega_i:=\omega_i-\omega_{0i}.
\label{eq:def_cOmega}
\end{equation}
By construction,
\begin{equation}
\int c_i f\,\dd\Gamma=0,
\qquad
\int \Omega_i f\,\dd\Gamma=0.
\label{eq:zero_mean}
\end{equation}
From this point onward we restrict attention to isotropic microinertia,
\begin{equation}
\bbI=I\,\Id,
\qquad
J:=\frac{I}{m},
\end{equation}
which covers perfectly rough spheres and isotropically distributed rigid
rotators.

For the local quasi-equilibrium introduced later we use a single temperature
$T$. The full heat-flux problem is not developed in this paper; the focus is on
the mechanical and spin transport channels.

\subsection{Collision invariants}
For elastic binary collisions the standard collision invariants are the particle
number, linear momentum, total angular momentum, and total kinetic energy
\cite{ChapmanCowling1970,Cercignani1988}. In particular we assume that
\begin{align}
\int C[f,f]\,\dd\Gamma&=0,
\label{eq:inv_mass}\\
\int mv_i\,C[f,f]\,\dd\Gamma&=0,
\label{eq:inv_momentum}\\
\int \Bigl[\eps_{ijk}x_jmv_k+(\bbI\bm{\omega})_i\Bigr]C[f,f]\,\dd\Gamma&=0.
\label{eq:inv_angmom}
\end{align}
The total-energy invariant will not be used explicitly below and is therefore
omitted from the displayed list. Equations
\eqref{eq:inv_mass}--\eqref{eq:inv_angmom} are sufficient for the
one-particle moment identities derived below. For finite-size rough particles,
however, the exact local coarse-grained momentum and angular-momentum balances
contain additional contact-transfer fluxes. In particular, the antisymmetric
part of the total stress produces a local intrinsic-spin torque. We therefore
distinguish below between the one-particle kinetic moments obtained directly
from Eq.~\eqref{eq:BC} and the exact finite-size balance laws used for the
retained-spin micropolar closure.

\section{Exact balance laws from the kinetic equation}
\subsection{Mass balance}
Integrating Eq.~\eqref{eq:BC} over phase space gives
\begin{equation}
\int \partial_t f\,\dd\Gamma
+
\int v_j\partial_{x_j}f\,\dd\Gamma
+
\int \frac{F^{\mathrm{ext}}_j}{m}\partial_{v_j}f\,\dd\Gamma
+
\int (\bbI^{-1}\tauext)_j\partial_{\omega_j}f\,\dd\Gamma
+
\int \mathcal{R}[f]\,\dd\Gamma
=
\int C[f,f]\,\dd\Gamma.
\end{equation}
Each term is evaluated separately.
\begin{align}
\int \partial_t f\,\dd\Gamma
&=\partial_t\int f\,\dd\Gamma
=\partial_t n,
\\
\int v_j\partial_{x_j}f\,\dd\Gamma
&=\partial_{x_j}\int v_j f\,\dd\Gamma
=\partial_{x_j}(nu_j),
\\
\int \frac{F^{\mathrm{ext}}_j}{m}\partial_{v_j}f\,\dd\Gamma
&=\frac{F^{\mathrm{ext}}_j}{m}\int \partial_{v_j}f\,\dd\Gamma
=0,
\\
\int (\bbI^{-1}\tauext)_j\partial_{\omega_j}f\,\dd\Gamma
&=(\bbI^{-1}\tauext)_j\int \partial_{\omega_j}f\,\dd\Gamma
=0,
\\
\int \mathcal{R}[f]\,\dd\Gamma&=0,
\\
\int C[f,f]\,\dd\Gamma&=0.
\end{align}
Thus
\begin{equation}
\partial_t n+\partial_{x_j}(nu_j)=0.
\label{eq:continuity_n}
\end{equation}
Multiplying by $m$ gives the continuity equation
\begin{equation}
\partial_t\rho+\Div(\rho\bm{u})=0.
\label{eq:continuity_rho}
\end{equation}

\subsection{Linear-momentum balance}
Multiply Eq.~\eqref{eq:BC} by $mv_i$ and integrate over phase space. The time
and transport terms are
\begin{align}
\int mv_i\partial_t f\,\dd\Gamma
&=\partial_t\int mv_i f\,\dd\Gamma
=\partial_t(\rho u_i),
\label{eq:mom_time}\\
\int mv_i v_j\partial_{x_j}f\,\dd\Gamma
&=\partial_{x_j}\int mv_i v_j f\,\dd\Gamma.
\label{eq:mom_transport}
\end{align}
For the force term we integrate by parts in velocity:
\begin{align}
\int mv_i\frac{F^{\mathrm{ext}}_j}{m}\partial_{v_j}f\,\dd\Gamma
&=F^{\mathrm{ext}}_j\int v_i\partial_{v_j}f\,\dd\Gamma
\nonumber\\
&=F^{\mathrm{ext}}_j\int\Bigl[\partial_{v_j}(v_if)-\delta_{ij}f\Bigr]\,\dd\Gamma
\nonumber\\
&=-F^{\mathrm{ext}}_i\int f\,\dd\Gamma
=-nF^{\mathrm{ext}}_i.
\label{eq:mom_force}
\end{align}
The torque term vanishes because $v_i$ is independent of $\bm{\omega}$:
\begin{equation}
\int mv_i(\bbI^{-1}\tauext)_j\partial_{\omega_j}f\,\dd\Gamma=0.
\end{equation}
The orientation-streaming term also vanishes,
\begin{equation}
\int mv_i\mathcal{R}[f]\,\dd\Gamma=0,
\end{equation}
and the collision term vanishes by Eq.~\eqref{eq:inv_momentum}:
\begin{equation}
\int mv_i C[f,f]\,\dd\Gamma=0.
\end{equation}

We therefore obtain
\begin{equation}
\partial_t(\rho u_i)+\partial_{x_j}\Pi_{ij}^{(k)}=\rho F_i,
\qquad
F_i:=\frac{F_i^{\mathrm{ext}}}{m},
\label{eq:mom_balance_raw}
\end{equation}
where the kinetic momentum-flux tensor is
\begin{equation}
\Pi_{ij}^{(k)}:=\int mv_iv_jf\,\dd\Gamma.
\label{eq:def_Pik}
\end{equation}
To separate convection from stress, write $v_i=u_i+c_i$ and expand:
\begin{align}
\Pi_{ij}^{(k)}
&=\int m(u_i+c_i)(u_j+c_j)f\,\dd\Gamma
\nonumber\\
&=mu_iu_j\int f\,\dd\Gamma
+mu_i\int c_jf\,\dd\Gamma
+mu_j\int c_if\,\dd\Gamma
+\int mc_ic_jf\,\dd\Gamma.
\end{align}
Using Eq.~\eqref{eq:zero_mean} gives
\begin{equation}
\Pi_{ij}^{(k)}=\rho u_i u_j+\int mc_ic_jf\,\dd\Gamma.
\label{eq:Pik_split}
\end{equation}
The kinetic Cauchy stress is therefore
\begin{equation}
\sigma_{ij}^{(k)}:=-\int mc_ic_jf\,\dd\Gamma.
\label{eq:def_sigma_k}
\end{equation}
Because $c_ic_j$ is symmetric, $\sigma^{(k)}_{ij}$ is symmetric. Equation
\eqref{eq:mom_balance_raw} may then be written as
\begin{equation}
\partial_t(\rho u_i)+\partial_{x_j}(\rho u_iu_j-\sigma_{ij}^{(k)})=\rho F_i,
\end{equation}
or, equivalently,
\begin{equation}
\rho\frac{\D u_i}{\D t}=\partial_{x_j}\sigma^{(k)}_{ij}+\rho F_i.
\label{eq:mom_balance_exact_k}
\end{equation}

For structured or finite-size particles, Eq.~\eqref{eq:mom_balance_exact_k}
identifies only the one-particle kinetic contribution. The exact coarse-grained
local momentum balance contains an additional contact-transfer stress
$\sigma_{ij}^{(\intc)}$. We therefore reserve the notation
\begin{equation}
\sigma_{ij}:=\sigma_{ij}^{(k)}+\sigma_{ij}^{(\intc)}
\label{eq:def_total_sigma}
\end{equation}
for the \emph{total} Cauchy stress and define the corresponding total momentum
flux by
\begin{equation}
\Pi_{ij}:=\rho u_i u_j-\sigma_{ij}
=\Pi_{ij}^{(k)}-\sigma_{ij}^{(\intc)}.
\label{eq:def_Pi_total}
\end{equation}
The exact finite-size momentum balance is then
\begin{equation}
\partial_t(\rho u_i)+\partial_{x_j}\Pi_{ij}=\rho F_i,
\qquad
\rho\frac{\D u_i}{\D t}=\partial_{x_j}\sigma_{ij}+\rho F_i.
\label{eq:mom_balance_exact_total}
\end{equation}
In a point-particle theory $\sigma_{ij}^{(\intc)}\equiv 0$ and
Eq.~\eqref{eq:mom_balance_exact_total} reduces to
Eq.~\eqref{eq:mom_balance_exact_k}.

\subsection{Total and intrinsic angular momentum}
The exact intrinsic-spin balance is obtained most cleanly by first writing the
balance of \emph{total} angular momentum and then subtracting the orbital part.
For a single particle, the total angular momentum about the origin is
\begin{equation}
\ell_i:=\eps_{ijk}x_jmv_k+(\bbI\bm{\omega})_i.
\label{eq:def_ell}
\end{equation}
Multiply Eq.~\eqref{eq:BC} by $\ell_i$ and integrate over phase space.

The time-derivative term gives
\begin{equation}
\int \ell_i\partial_tf\,\dd\Gamma=\partial_t\int \ell_if\,\dd\Gamma.
\label{eq:ang_time}
\end{equation}
For the transport term we must differentiate $\ell_i$ with respect to
$\bm{x}$ because $x_j$ is not integrated over:
\begin{align}
\int \ell_i v_j\partial_{x_j}f\,\dd\Gamma
&=\partial_{x_j}\int \ell_iv_jf\,\dd\Gamma
-\int (\partial_{x_j}\ell_i)v_jf\,\dd\Gamma.
\label{eq:ang_transport1}
\end{align}
Now
\begin{equation}
\partial_{x_j}\ell_i
=\partial_{x_j}(\eps_{ipq}x_pmv_q)
=\eps_{ipq}\delta_{jp}mv_q
=\eps_{ijq}mv_q,
\label{eq:grad_ell}
\end{equation}
so Eq.~\eqref{eq:ang_transport1} becomes
\begin{equation}
\int \ell_i v_j\partial_{x_j}f\,\dd\Gamma
=\partial_{x_j}\int \ell_iv_jf\,\dd\Gamma
-\eps_{ijq}\int mv_qv_jf\,\dd\Gamma.
\label{eq:ang_transport2}
\end{equation}

The force term gives the expected external moment:
\begin{align}
\int \ell_i\frac{F_j^{\mathrm{ext}}}{m}\partial_{v_j}f\,\dd\Gamma
&=-\int \partial_{v_j}\ell_i\,\frac{F_j^{\mathrm{ext}}}{m}f\,\dd\Gamma
\nonumber\\
&=-\eps_{ipq}x_p F_q^{\mathrm{ext}}\int f\,\dd\Gamma
\nonumber\\
&=-\rho(\bm{x}\times\bm{F})_i.
\label{eq:ang_force}
\end{align}
The torque term is
\begin{align}
\int \ell_i(\bbI^{-1}\tauext)_j\partial_{\omega_j}f\,\dd\Gamma
&=-\int \partial_{\omega_j}\ell_i\,(\bbI^{-1}\tauext)_j f\,\dd\Gamma
\nonumber\\
&=-\tau_i^{\mathrm{ext}}\int f\,\dd\Gamma
=-n\tau_i^{\mathrm{ext}}.
\label{eq:ang_torque}
\end{align}
The orientation-streaming term integrates to zero, and the collision term
vanishes because $\ell_i$ is a collision invariant:
\begin{equation}
\int \ell_iC[f,f]\,\dd\Gamma=0.
\end{equation}
Collecting terms yields the balance of total angular momentum,
\begin{equation}
\partial_t\mathcal{L}_i
+\partial_{x_j}\mathcal{J}_{ij}
=\eps_{ijq}\Pi^{(k)}_{qj}+\rho(\bm{x}\times\bm{F})_i+n\tau_i^{\mathrm{ext}},
\label{eq:total_ang_balance_raw}
\end{equation}
with
\begin{equation}
\mathcal{L}_i:=\int \ell_i f\,\dd\Gamma,
\qquad
\mathcal{J}_{ij}:=\int \ell_i v_jf\,\dd\Gamma.
\end{equation}

Now split total angular momentum into orbital and intrinsic parts. Since
\begin{equation}
\int mv_qf\,\dd\Gamma=\rho u_q,
\qquad
\int (\bbI\bm{\omega})_if\,\dd\Gamma=In\omega_{0i}=\rho J\omega_{0i},
\end{equation}
we have
\begin{equation}
\mathcal{L}_i=\eps_{ipq}x_p\rho u_q+\rho J\omega_{0i}.
\label{eq:L_split}
\end{equation}
Likewise, the one-particle part of the angular-momentum flux is
\begin{align}
\mathcal{J}^{(k)}_{ij}
&=\eps_{ipq}x_p\Pi^{(k)}_{qj}
+\int (\bbI\bm{\omega})_i v_jf\,\dd\Gamma
\nonumber\\
&=\eps_{ipq}x_p\Pi^{(k)}_{qj}+\rho J\omega_{0i}u_j-m_{ij}^{(k)},
\label{eq:J_split}
\end{align}
where the kinetic couple stress is defined as minus the nonconvective intrinsic
flux
\begin{equation}
m_{ij}^{(k)}:=-\int (\bbI\bm{\omega})_i(v_j-u_j)f\,\dd\Gamma
=-\int (\bbI\bm{\omega})_ic_jf\,\dd\Gamma.
\label{eq:def_mij}
\end{equation}
For finite-size particles the exact coarse-grained total angular-momentum flux
contains additional contact-transfer contributions. We therefore write
\begin{equation}
\mathcal{J}_{ij}:=\eps_{ipq}x_p\Pi_{qj}+\rho J\omega_{0i}u_j-m_{ij},
\label{eq:def_total_Jij}
\end{equation}
where $m_{ij}$ is the exact intrinsic flux. In the dilute first-order closure
developed below, its leading contribution is represented by the one-particle
moment $m_{ij}^{(k)}$.

Next derive the balance of orbital angular momentum directly from the exact
finite-size momentum equation \eqref{eq:mom_balance_exact_total}. Multiplying by
$\eps_{ipq}x_p$ gives
\begin{align}
\partial_t(\eps_{ipq}x_p\rho u_q)
+\eps_{ipq}x_p\partial_{x_j}\Pi_{qj}
&=\rho(\bm{x}\times\bm{F})_i.
\end{align}
Using the product rule,
\begin{equation}
\eps_{ipq}x_p\partial_{x_j}\Pi_{qj}
=\partial_{x_j}(\eps_{ipq}x_p\Pi_{qj})-\eps_{ijq}\Pi_{qj},
\end{equation}
so the orbital balance is
\begin{equation}
\partial_t(\eps_{ipq}x_p\rho u_q)
+\partial_{x_j}(\eps_{ipq}x_p\Pi_{qj})
=\rho(\bm{x}\times\bm{F})_i+\eps_{ijq}\Pi_{qj}.
\label{eq:orbital_balance}
\end{equation}
The corresponding exact coarse-grained balance of total angular momentum reads
\begin{equation}
\partial_t\mathcal{L}_i
+\partial_{x_j}\mathcal{J}_{ij}
=\rho(\bm{x}\times\bm{F})_i+\rho G_i,
\label{eq:total_ang_balance_exact}
\end{equation}
where, as before,
\begin{equation}
\rho G_i:=n\tau_i^{\mathrm{ext}}.
\end{equation}
Subtracting Eq.~\eqref{eq:orbital_balance} from
Eq.~\eqref{eq:total_ang_balance_exact} and using
Eqs.~\eqref{eq:L_split} and \eqref{eq:def_total_Jij} gives the exact
finite-size intrinsic-spin balance
\begin{equation}
\partial_t(\rho J\omega_{0i})
+\partial_{x_j}(\rho J\omega_{0i}u_j-m_{ij})
=-\eps_{ijq}\Pi_{qj}+\rho G_i.
\label{eq:spin_balance_conservative}
\end{equation}
Since the convective part $\rho u_q u_j$ is symmetric,
$-\eps_{ijq}\Pi_{qj}=-\eps_{ijk}\sigma_{jk}$, and we may write
\begin{equation}
\partial_t(\rho J\bm{\omega}_0)+\Div(\rho J\bm{\omega}_0\otimes\bm{u}-\bm{m})
=\bm{\tau}^{(\sigma)}+\rho\bm{G},
\qquad
\tau_i^{(\sigma)}:=-\eps_{ijk}\sigma_{jk}.
\label{eq:spin_balance_exact}
\end{equation}
Using the continuity equation in the usual way gives the nonconservative form
\begin{equation}
\rho J\frac{\D\omega_{0i}}{\D t}
=\partial_{x_j}m_{ij}-\eps_{ijk}\sigma_{jk}+\rho G_i.
\label{eq:spin_balance_exact_nc}
\end{equation}

\subsection{Where does the antisymmetric stress appear?}
Equation \eqref{eq:spin_balance_exact_nc} already contains the
antisymmetric-stress torque through the exact total stress. In a point-particle
theory $\sigma_{ij}=\sigma_{ij}^{(k)}$ is symmetric, so
$\tau_i^{(\sigma)}=0$ and the intrinsic balance reduces to the purely kinetic
flux form. For finite-size rough particles, however, the contact-transfer part
$\sigma_{ij}^{(\intc)}$ need not be symmetric, and
\begin{equation}
\tau_i^{(\sigma)}:=-\eps_{ijk}\sigma_{jk}
=-\eps_{ijk}\sigma_{jk}^{(\intc)}
\label{eq:spin_balance_sigma_torque}
\end{equation}
represents the local orbital--intrinsic exchange. This is the exact channel that
later carries the rotational viscosity $\eta_r$; it is not an additional
constitutive term to be appended after the fact. We return to this point in
Sec.~\ref{sec:first_order_constitutive}.

\HLTXT_BLUE{The connection between the microscopic collision torque and $\tau_i^{(\sigma)}$ is therefore a coarse-grained finite-size statement. A binary rough-sphere collision applies equal and opposite impulses at contact points that are displaced from the particle centers. The same impulse changes the orbital angular momentum of the pair and the intrinsic spins. In a local balance law, the transfer of linear momentum across a coarse-graining surface appears as a contact-transfer stress, while the moment of the same transfer appears as an orbital--intrinsic exchange. Requiring local conservation of total angular momentum for arbitrary control volumes converts this exchange into the antisymmetric part of the total Cauchy stress. Thus $-\epsilon_{ijk}\sigma_{jk}$ is not a new body torque and is not produced by the symmetric one-particle kinetic stress; it is the coarse-grained representation of the finite-size contact-transfer torque.}

\section{Generalized Chapman--Enskog construction with retained spin}
\subsection{Scaling and extended-hydrodynamic viewpoint}
Let $L$ be a macroscopic length scale and $\ell$ a mean free path. The Knudsen
number is
\begin{equation}
\eps:=\frac{\ell}{L}\ll 1.
\end{equation}
The Chapman--Enskog method assumes that the distribution function depends on
$(\bm{x},t)$ only through slowly varying macroscopic fields and expands
\begin{equation}
f=f^{(0)}+\eps f^{(1)}+\eps^2f^{(2)}+\cdots,
\qquad
\partial_t=\partial_t^{(0)}+\eps\partial_t^{(1)}+\cdots.
\label{eq:CE_expansion}
\end{equation}
The retained-spin closure uses the fields
\begin{equation}
(n,\bm{u},T,\bm{\omega}_0)
\end{equation}
as the coordinates of the reference manifold.

Strictly speaking, only the collision invariants are guaranteed to remain slow
in the ordinary dilute-gas hydrodynamic limit. The mean spin is generally not a
strict invariant and relaxes on a collisional time scale. Retaining
$\bm{\omega}_0$ therefore corresponds to a generalized or extended-hydrodynamic
description \cite{MullerRuggeri1998,JouCasasLebon1999}. This viewpoint is not a
mathematical defect but the intended regime: we want a closure that keeps the
local spin explicit rather than eliminating it instantaneously.

\HLTXT_BLUE{It is important not to confuse the retained-spin ordering with an assertion that $\bm{\omega}_0$ is an ordinary hydrodynamic invariant. For a generic rough-particle collision operator the homogeneous mean spin relaxes on the collisional time scale. In the notation used below, this corresponds to an $O(1)$ microscopic relaxation rate before any extended-hydrodynamic ordering is imposed. A strict hydrodynamic limit would therefore eliminate $\bm{\omega}_0$ and leave only an enslaved spin response. The present paper instead adopts a retained-spin or extended-hydrodynamic ordering in which the residual axial relaxation on the retained-spin manifold is counted at the same formal order as the first gradient corrections. This is a constitutive assumption about the regime being described, not a theorem following from the ordinary collision invariants.}

\subsection{Local quasi-equilibrium distribution}
The natural reference state for the retained-spin theory is the
maximum-entropy distribution consistent with $(n,\bm{u},T,\bm{\omega}_0)$,
namely
\begin{equation}
f^{(0)}
=
n
\left(\frac{m}{2\pi k_BT}\right)^{3/2}
\left(\frac{I}{2\pi k_BT}\right)^{3/2}
\exp\!\left[-\frac{mc^2}{2k_BT}-\frac{I\Omega^2}{2k_BT}\right].
\label{eq:f0}
\end{equation}
For general isotropic rotators the second prefactor becomes
$(\det\bbI)^{1/2}/(2\pi k_BT)^{3/2}$, but the scalar form \eqref{eq:f0} is
sufficient for the present paper.

The zeroth-order moments are immediate. The kinetic stress is isotropic,
\begin{equation}
\sigma_{ij}^{(k,0)}=-\int mc_ic_jf^{(0)}\,\dd\Gamma=-P\,\delta_{ij},
\qquad
P=nk_BT,
\label{eq:sigma0}
\end{equation}
and the couple stress vanishes,
\begin{equation}
m_{ij}^{(k,0)}=-\int I\omega_ic_jf^{(0)}\,\dd\Gamma=0,
\label{eq:m0}
\end{equation}
because $c_j$ has zero mean under the local Maxwellian. Viscous and
spin-diffusive effects therefore arise only at first order.

At this point the extended-hydrodynamic ordering should be stated explicitly.
For a strict classical Chapman--Enskog construction, the local Maxwellian lies
on the exact null manifold of the one-particle collision operator, so that
$C[f^{(0)},f^{(0)}]=0$. In the retained-spin closure for finite-size rough
particles, however, the axial collisional-transfer channel associated with the
exact spin torque need not vanish identically on the retained-spin manifold. We
therefore introduce an auxiliary bookkeeping parameter $\delta_s$ that measures
the residual axial relaxation strength and write the matched-order axial source
symbolically as
\begin{equation*}
\delta_s=O(\eps),
\qquad
\mathcal{S}^{(\ax)}_i a_i
=
\mathcal{S}^{(\ax,\nabla)}_i a_i
+\delta_s\,\mathcal{R}^{(\ax)}[f^{(0)}].
\tag{61a}
\end{equation*}
Here $\mathcal{R}^{(\ax)}[f^{(0)}]$ denotes the residual axial contribution of
the retained-spin manifold, understood to belong to the collisional-transfer
channel rather than to the symmetric one-particle kinetic stress. It is
orthogonal to the strict mass, linear-momentum, and energy invariants,
\begin{equation*}
\int \mathcal{R}^{(\ax)}[f^{(0)}]\,\dd\Gamma=0,
\qquad
\int mv_i\,\mathcal{R}^{(\ax)}[f^{(0)}]\,\dd\Gamma=0,
\qquad
\int \left(\frac{m}{2}v^2+\frac{I}{2}\omega^2\right)
\mathcal{R}^{(\ax)}[f^{(0)}]\,\dd\Gamma=0.
\tag{61b}
\end{equation*}
Operationally, the present retained-spin closure may therefore be regarded as a
double expansion in $\eps$ and $\delta_s$, with the bookkeeping
$\delta_s\sim\eps$. Under this matched ordering, the residual axial relaxation
enters at the same asymptotic order as the usual gradient-driven first-order
source. This is the precise sense in which $\bm{\omega}_0$ is treated as a
quasi-slow variable in the present paper. If instead $\delta_s=O(1)$, the
irreducible sector decomposition derived below still identifies the correct
constitutive channels, but the construction should then be read as a formal
extended-hydrodynamic closure rather than as a strict one-parameter
Chapman--Enskog limit. The explicit rough-sphere estimates derived later are
therefore coefficient evaluations within the same retained-spin constitutive
structure; they are not, by themselves, an additional proof of the ordering
$\delta_s\sim\eps$.

\HLTXT_BLUE{The explicit rough-sphere relaxation rate derived in Sec.~\ref{seq:perfectlyrhs} should therefore be read in the following way. It supplies the dilute-gas value of the coefficient that controls the axial relaxation channel once the retained-spin constitutive structure has been adopted. It does not prove that the retained-spin manifold is slow in the strict hydrodynamic sense. If the physical problem has $\delta_s=O(1)$ on the macroscopic time scale of interest, then the same algebra identifies the axial constitutive channel, but the spin variable should be eliminated rather than retained.}

\subsection{Linearized first-order equation}
Write the first-order correction in the standard multiplicative form
\begin{equation}
f^{(1)}=f^{(0)}\phi.
\label{eq:f1phi}
\end{equation}
The Chapman--Enskog matching conditions keep the retained fields entirely in
$f^{(0)}$; in particular,
\begin{equation}
\int f^{(1)}\,\dd\Gamma=0,
\qquad
\int c_i f^{(1)}\,\dd\Gamma=0,
\qquad
\int \Omega_i f^{(1)}\,\dd\Gamma=0.
\label{eq:CE_matching}
\end{equation}
Define the linearized collision operator around $f^{(0)}$ by
\begin{equation}
\mathcal{L}[\phi]
:=-\frac{1}{f^{(0)}}\Bigl(C[f^{(0)}\phi,f^{(0)}]+C[f^{(0)},f^{(0)}\phi]\Bigr).
\label{eq:linearized_operator}
\end{equation}
Then the first-order Chapman--Enskog equation is
\begin{equation}
\mathcal{L}[\phi]=\mathcal{S},
\label{eq:phi_equation}
\end{equation}
where the source $\mathcal{S}$ collects the $O(\eps)$ streaming contributions together with the matched-order residual axial collisional-transfer contribution described above.
The standard density, temperature, and heat-flux sectors follow the classical
Chapman--Enskog pattern \cite{ChapmanCowling1970,Sone2002}. Here we focus on
the sectors relevant to the stress and couple stress.

\subsection{Useful derivatives of the local Maxwellian}
The logarithm of \eqref{eq:f0} is
\begin{equation}
\ln f^{(0)}=
\ln n
-3\ln(2\pi k_BT)
+\frac{3}{2}\ln m+\frac{3}{2}\ln I
-\frac{mc^2}{2k_BT}
-\frac{I\Omega^2}{2k_BT}.
\end{equation}
At fixed $(\bm{v},\bm{\omega})$,
\begin{equation}
\partial_{x_\ell}c_i=-\partial_{x_\ell}u_i,
\qquad
\partial_{x_\ell}\Omega_i=-\partial_{x_\ell}\omega_{0i}.
\label{eq:grad_cOmega}
\end{equation}
Consequently,
\begin{align}
\partial_{x_\ell}c^2&=2c_i\partial_{x_\ell}c_i=-2c_i\partial_{x_\ell}u_i,
\\
\partial_{x_\ell}\Omega^2&=2\Omega_i\partial_{x_\ell}\Omega_i=-2\Omega_i\partial_{x_\ell}\omega_{0i}.
\end{align}
Differentiating $\ln f^{(0)}$ with respect to $x_\ell$ gives
\begin{align}
\partial_{x_\ell}\ln f^{(0)}
&=\partial_{x_\ell}\ln n
-3\partial_{x_\ell}\ln T
+\frac{mc^2+I\Omega^2}{2k_BT}\partial_{x_\ell}\ln T
+\frac{m}{k_BT}c_i\partial_{x_\ell}u_i
+\frac{I}{k_BT}\Omega_i\partial_{x_\ell}\omega_{0i}
\nonumber\\
&=\partial_{x_\ell}\ln n
+\left[\frac{mc^2+I\Omega^2}{2k_BT}-3\right]\partial_{x_\ell}\ln T
+\frac{m}{k_BT}c_i\partial_{x_\ell}u_i
+\frac{I}{k_BT}\Omega_i\partial_{x_\ell}\omega_{0i}.
\label{eq:grad_logf0}
\end{align}
This is the basic starting point for the first-order source decomposition.

The full source $\mathcal{S}$ also contains Euler-level time derivatives of the
slow fields. After the usual elimination of the invariant channels by the
zeroth-order balance equations, the mechanical and spin-gradient part of the
source takes the schematic form
\begin{equation}
\mathcal{S}_{\mech}
=-\frac{m}{k_BT}c_ic_j\partial_j u_i
-\frac{I}{k_BT}c_j\Omega_i\partial_j\omega_{0i}
+\mathcal{S}^{(\ax)}_i a_i
+\cdots,
\label{eq:Smech_raw}
\end{equation}
where the dots denote density, temperature, and heat-flux channels, and
\begin{equation}
a_i:=\frac12\zeta_i-\omega_{0i},
\qquad
\zeta_i:=(\Curl\bm{u})_i.
\label{eq:def_ai}
\end{equation}
The last term $\mathcal{S}^{(\ax)}_i a_i$ is written separately because the
axial mismatch sector is not generated by the symmetric one-particle kinetic
stress alone. In the matched ordering described above, this symbol also absorbs
the $O(\eps)$ residual axial relaxation of the retained-spin
quasi-equilibrium manifold. The axial sector therefore belongs to the
collisional-transfer channel and will be treated formally in
Sec.~\ref{sec:formal_coefficient_problem}.

\subsection{Irreducible decomposition of the velocity and spin gradients}
Introduce the standard decomposition of the velocity gradient,
\begin{equation}
\partial_i u_j=D_{ij}+\frac13\theta\,\delta_{ij}+W_{ij},
\qquad
\theta:=\partial_k u_k,
\qquad
W_{ij}:=\frac12(\partial_i u_j-\partial_j u_i),
\label{eq:grad_u_decomp}
\end{equation}
where $D_{ij}$ is symmetric traceless:
\begin{equation}
D_{ij}=\frac12(\partial_i u_j+\partial_j u_i)-\frac13\theta\,\delta_{ij}.
\end{equation}
Because $c_ic_j$ is symmetric,
\begin{equation}
c_ic_jW_{ij}=0.
\label{eq:ci_cj_W_zero}
\end{equation}
Therefore
\begin{equation}
c_ic_j\partial_j u_i
=\STF{c_ic_j}D_{ij}+\frac13c^2\theta.
\label{eq:velocity_source_split}
\end{equation}
Equation \eqref{eq:velocity_source_split} is crucial: the one-particle kinetic
streaming source contains \emph{no} axial contribution proportional to the
antisymmetric velocity gradient.

Now decompose the spin gradient into irreducible parts. Define
\begin{equation}
q:=\partial_k\omega_{0k},
\qquad
b_i:=\frac12\eps_{ijk}\partial_j\omega_{0k}=\frac12(\Curl\bm{\omega}_0)_i,
\label{eq:def_qb}
\end{equation}
and the symmetric traceless tensor
\begin{equation}
E_{ij}:=\frac12(\partial_i\omega_{0j}+\partial_j\omega_{0i})-\frac13q\,\delta_{ij}.
\label{eq:def_Eij}
\end{equation}
Then
\begin{equation}
\partial_i\omega_{0j}=E_{ij}+\frac13q\,\delta_{ij}+\eps_{ijk}b_k.
\label{eq:grad_omega_decomp}
\end{equation}
Using this in the second term of Eq.~\eqref{eq:Smech_raw} gives
\begin{align}
c_j\Omega_i\partial_j\omega_{0i}
&=c_j\Omega_i\left(E_{ji}+\frac13q\,\delta_{ji}+\eps_{jik}b_k\right)
\nonumber\\
&=\STF{c_j\Omega_i}E_{ij}+\frac13(\bm{c}\cdot\bm{\Omega})q+(\bm{c}\times\bm{\Omega})_kb_k.
\label{eq:spin_source_split}
\end{align}
The natural basis functions in the spin-gradient sector are therefore
\begin{equation}
Y^{(0)}:=\bm{c}\cdot\bm{\Omega},
\qquad
Y_i^{(1)}:=(\bm{c}\times\bm{\Omega})_i,
\qquad
Y_{ij}^{(2)}:=\STF{c_j \Omega_i }.
\label{eq:Y_basis}
\end{equation}

\section{First-order constitutive structure}
\label{sec:first_order_constitutive}
\subsection{Sectorwise expansion of the first-order correction}
Rotational invariance implies that the first-order correction can be expanded in
irreducible sectors. Restricting attention to the channels relevant for the
stress and couple stress, we write
\begin{equation}
\phi
=
B^{(\dev)}_{ij}D_{ij}
+B^{(\tr)}\theta
+X_i a_i
+Y\,q
+Z_ib_i
+Z_{ij}E_{ij}
+\cdots,
\label{eq:phi_sectorwise}
\end{equation}
where $Z_{ij}$ is symmetric traceless in $(i,j)$ and the omitted terms belong
to heat and diffusion sectors. The response functions
$B^{(\dev)}_{ij},B^{(\tr)},X_i,Y,Z_i,Z_{ij}$ are functions of the microscopic
variables $(\bm{c},\bm{\Omega})$.

Because the linearized collision operator commutes with proper rotations in the
isotropic case, the distinct irreducible sectors decouple at first order. This
is the basic reason why one may speak of \emph{the} shear channel, \emph{the}
bulk channel, \emph{the} spin-diffusion channels, and \emph{the} axial mismatch
channel.

\subsection{Total stress at first order}
The first-order kinetic stress is
\begin{equation}
\sigma_{ij}^{(k,1)}=-\int mc_ic_jf^{(0)}\phi\,\dd\Gamma.
\label{eq:sigma_k_firstorder}
\end{equation}
The kernel $c_ic_jf^{(0)}$ is symmetric in $(i,j)$ and even under
$\bm{\Omega}\mapsto-\bm{\Omega}$. The axial-mismatch and spin-gradient sectors in
Eq.~\eqref{eq:phi_sectorwise} are odd in $\bm{\Omega}$ and therefore integrate
to zero against this kernel; in particular, no $q\,\delta_{ij}$ or $E_{ij}$
contribution survives in the kinetic stress. Thus only the symmetric
velocity-gradient sectors contribute to $\sigma_{ij}^{(k,1)}$, and rotational
invariance forces the result to be of the form
\begin{equation}
\sigma_{ij}^{(k)}
=-P\,\delta_{ij}+2\eta D_{ij}+\xi\theta\,\delta_{ij}.
\label{eq:sigma_kinetic_const}
\end{equation}
The coefficients $\eta$ and $\xi$ are the shear and bulk viscosities.

The antisymmetric channel belongs to the collisional-transfer part of the total stress. Introduce the axial mismatch tensor
\begin{equation}
A_{ij}:=W_{ij}-\eps_{ijk}\omega_{0k}.
\label{eq:def_Aij}
\end{equation}
Since
\begin{equation}
W_{ij}=\frac12\eps_{ijk}\zeta_k,
\end{equation}
we can also write
\begin{equation}
A_{ij}=\eps_{ijk}a_k,
\qquad
a_k=\frac12\zeta_k-\omega_{0k}.
\end{equation}
Isotropy and parity imply that the first-order intrinsic stress in this
channel must itself be antisymmetric and proportional to $A_{ij}$, so we
define
\begin{equation}
\sigma_{ij}^{(\intc,1)}=-2\eta_rA_{ij}.
\label{eq:sigma_axial_def}
\end{equation}
Equivalently,
\begin{equation}
\sigma_{ij}^{(\intc,1)}
=\eta_r\eps_{ijk}(2\omega_{0k}-\zeta_k),
\qquad
-\eps_{ijk}\sigma_{jk}^{(\intc,1)}=2\eta_r(\zeta_i-2\omega_{0i}).
\label{eq:sigma_axial_epsform}
\end{equation}
Combining Eqs.~\eqref{eq:sigma_kinetic_const} and \eqref{eq:sigma_axial_def},
the total first-order stress is
\begin{equation}
\sigma_{ij}
=-P\,\delta_{ij}
+2\eta D_{ij}
+\xi\theta\,\delta_{ij}
-2\eta_rW_{ij}
+2\eta_r\eps_{ijk}\omega_{0k}.
\label{eq:sigma_total_constitutive}
\end{equation}
This is the standard micropolar stress with the present sign convention.

\subsection{Divergence of the total stress}
We now compute $\partial_j\sigma_{ij}$ explicitly because the intermediate
vector identities are often skipped.
Using Eq.~\eqref{eq:sigma_total_constitutive},
\begin{align}
\partial_j\sigma_{ij}
&=-\partial_iP
+2\eta\,\partial_jD_{ij}
+\xi\,\partial_i\theta
-2\eta_r\,\partial_jW_{ij}
+2\eta_r\,\partial_j(\eps_{ijk}\omega_{0k}).
\label{eq:divsigma_step1}
\end{align}
The divergence of the symmetric traceless rate is
\begin{align}
\partial_jD_{ij}
&=\partial_j\left[\frac12(\partial_iu_j+\partial_ju_i)-\frac13\theta\,\delta_{ij}\right]
\nonumber\\
&=\frac12\partial_i\partial_ju_j+\frac12\partial_j\partial_ju_i-\frac13\partial_i\theta
\nonumber\\
&=\frac12\partial_i\theta+\frac12\grad^2u_i-\frac13\partial_i\theta
\nonumber\\
&=\frac12\grad^2u_i+\frac16\partial_i\theta.
\label{eq:divD}
\end{align}
Hence
\begin{equation}
2\eta\,\partial_jD_{ij}=\eta\grad^2u_i+\frac{\eta}{3}\partial_i\theta.
\label{eq:2eta_divD}
\end{equation}
For the antisymmetric velocity-gradient term,
\begin{align}
-2\eta_r\partial_jW_{ij}
&=-\eta_r\partial_j(\partial_iu_j-\partial_ju_i)
\nonumber\\
&=-\eta_r(\partial_i\theta-\grad^2u_i)
\nonumber\\
&=\eta_r\grad^2u_i-\eta_r\partial_i\theta.
\label{eq:divWterm}
\end{align}
Finally,
\begin{equation}
2\eta_r\partial_j(\eps_{ijk}\omega_{0k})=2\eta_r(\Curl\bm{\omega}_0)_i.
\label{eq:curlomegaterm}
\end{equation}
Substituting Eqs.~\eqref{eq:2eta_divD}, \eqref{eq:divWterm}, and
\eqref{eq:curlomegaterm} into Eq.~\eqref{eq:divsigma_step1} gives
\begin{equation}
\partial_j\sigma_{ij}
=-\partial_iP
+(\eta+\eta_r)\grad^2u_i
+\left(\frac{\eta}{3}+\xi-\eta_r\right)\partial_i\theta
+2\eta_r(\Curl\bm{\omega}_0)_i.
\label{eq:divsigma_final}
\end{equation}
The momentum equation therefore becomes
\begin{equation}
\rho\frac{\D u_i}{\D t}
=-\partial_iP
+(\eta+\eta_r)\grad^2u_i
+\left(\frac{\eta}{3}+\xi-\eta_r\right)\partial_i\theta
+2\eta_r(\Curl\bm{\omega}_0)_i
+\rho F_i.
\label{eq:momentum_micropolar}
\end{equation}

\subsection{Couple stress at first order} \label{subsec:couplestressatfirstord}
The leading one-particle contribution to the first-order couple stress is
obtained from Eq.~\eqref{eq:def_mij}:
\begin{equation}
m_{ij}^{(k,1)}=-\int I\omega_ic_jf^{(0)}\phi\,\dd\Gamma.
\end{equation}
Writing $\omega_i=\Omega_i+\omega_{0i}$ gives
\begin{equation}
m_{ij}^{(k,1)}
=-\int I\Omega_ic_jf^{(0)}\phi\,\dd\Gamma
-\omega_{0i}\int Ic_jf^{(0)}\phi\,\dd\Gamma.
\end{equation}
The second term vanishes by the Chapman--Enskog matching condition
$\int c_jf^{(1)}\,\dd\Gamma=\int c_jf^{(0)}\phi\,\dd\Gamma=0$ from
Eq.~\eqref{eq:CE_matching}. Introduce the microscopic kernel
\begin{equation}
Q_{ij}:=I\Omega_ic_j.
\label{eq:def_Qij}
\end{equation}
Then
\begin{equation}
m_{ij}^{(k,1)}=-\int Q_{ij}f^{(0)}\phi\,\dd\Gamma.
\label{eq:moment_mij}
\end{equation}
At dilute first order we represent the total couple stress by this leading
one-particle contribution, so only the $q$, $b_i$, and $E_{ij}$ sectors
contribute. Rotational invariance therefore implies the general form
\begin{equation}
m_{ij}
=\lambda_0q\,\delta_{ij}+\lambda_1\eps_{ijk}b_k+\lambda_2E_{ij}.
\label{eq:mij_lambda_form}
\end{equation}
The coefficients are linear functionals of the three response functions
$Y,Z_i,Z_{ij}$. A convenient extraction is
\begin{align}
\lambda_0&=-\frac13\int Q_{kk}f^{(0)}Y\,\dd\Gamma,
\label{eq:lambda0_formula}\\
\lambda_1&=-\frac16\eps_{ijk}\int Q_{ij}f^{(0)}Z_k\,\dd\Gamma,
\label{eq:lambda1_formula}\\
\lambda_2&=-\frac15\int \STF{Q_{ij}}f^{(0)}Z_{ij}\,\dd\Gamma.
\label{eq:lambda2_formula}
\end{align}

To compare with the conventional micropolar notation, start from
\begin{equation}
m_{ij}
=\alpha\,\delta_{ij}\,\partial_k\omega_{0k}
+\beta(\partial_i\omega_{0j}+\partial_j\omega_{0i})
+\gamma(\partial_j\omega_{0i}-\partial_i\omega_{0j}).
\label{eq:mij_abg}
\end{equation}
Using Eq.~\eqref{eq:grad_omega_decomp},
\begin{align}
\partial_i\omega_{0j}+\partial_j\omega_{0i}
&=2E_{ij}+\frac{2}{3}q\,\delta_{ij},
\\
\partial_j\omega_{0i}-\partial_i\omega_{0j}
&=-2\eps_{ijk}b_k.
\end{align}
Substituting into Eq.~\eqref{eq:mij_abg} gives
\begin{equation}
m_{ij}=
\left(\alpha+\frac{2}{3}\beta\right)q\,\delta_{ij}
-2\gamma\,\eps_{ijk}b_k
+2\beta E_{ij}.
\label{eq:mij_abg_split}
\end{equation}
Comparing with Eq.~\eqref{eq:mij_lambda_form} yields
\begin{equation}
\lambda_0=\alpha+\frac23\beta,
\qquad
\lambda_1=-2\gamma,
\qquad
\lambda_2=2\beta.
\label{eq:lambda_to_abg}
\end{equation}
Equivalently,
\begin{equation}
\beta=\frac12\lambda_2,
\qquad
\gamma=-\frac12\lambda_1,
\qquad
\alpha=\lambda_0-\frac13\lambda_2.
\label{eq:abg_from_lambda}
\end{equation}

\subsection{Divergence of the couple stress}
Starting from Eq.~\eqref{eq:mij_abg},
\begin{align}
\partial_jm_{ij}
&=\alpha\,\partial_i\partial_k\omega_{0k}
+\beta\,\partial_j(\partial_i\omega_{0j}+\partial_j\omega_{0i})
+\gamma\,\partial_j(\partial_j\omega_{0i}-\partial_i\omega_{0j})
\nonumber\\
&=\alpha\,\partial_iq
+\beta(\partial_iq+\grad^2\omega_{0i})
+\gamma(\grad^2\omega_{0i}-\partial_iq)
\nonumber\\
&=(\beta+\gamma)\grad^2\omega_{0i}
+(\alpha+\beta-\gamma)\partial_iq.
\label{eq:divmij_final}
\end{align}
The exact spin balance \eqref{eq:spin_balance_exact_nc} therefore becomes,
after inserting the first-order constitutive forms for $m_{ij}$ and
$\sigma_{ij}$,
\begin{equation}
\rho J\frac{\D\omega_{0i}}{\D t}
=(\beta+\gamma)\grad^2\omega_{0i}
+(\alpha+\beta-\gamma)\partial_i(\Div\bm{\omega}_0)
+2\eta_r(\zeta_i-2\omega_{0i})
+\rho G_i.
\label{eq:spin_micropolar}
\end{equation}
Equations \eqref{eq:momentum_micropolar} and \eqref{eq:spin_micropolar} are the
retained-spin micropolar equations derived from the first-order constitutive
structure.

\section{Formal first-order coefficient problem}
\label{sec:formal_coefficient_problem}
The derivation above determines the tensorial \emph{form} of the constitutive
laws. We now spell out the first-order Chapman--Enskog problems that define the
coefficients themselves.

\subsection{Shear and bulk sectors}
The shear source in Eq.~\eqref{eq:Smech_raw} is proportional to
$\STF{c_ic_j}D_{ij}$, so the shear response satisfies a tensor equation of the
form
\begin{equation}
\mathcal{L}[B^{(\dev)}_{ij}]
=-\frac{m}{k_BT}\STF{c_ic_j}.
\label{eq:shear_integral_eq}
\end{equation}
The shear viscosity is then extracted from
\begin{equation}
\eta=-\frac{1}{10}\int m\STF{c_ic_j}f^{(0)}B^{(\dev)}_{ij}\,\dd\Gamma.
\label{eq:eta_formula}
\end{equation}
Likewise the scalar bulk channel may be written as
\begin{equation}
\mathcal{L}[B^{(\tr)}]=S^{(\tr)},
\label{eq:bulk_integral_eq}
\end{equation}
with the corresponding moment formula
\begin{equation}
\xi=-\frac13\int m\left(\frac13c^2-\frac{k_BT}{m}\right)f^{(0)}B^{(\tr)}\,\dd\Gamma.
\label{eq:xi_formula}
\end{equation}
The precise scalar basis in the bulk channel depends on how the invariant part
is projected out. This is standard and is not the focus of the present paper.

\subsection{Spin-diffusion sectors}
From Eqs.~\eqref{eq:Smech_raw} and \eqref{eq:spin_source_split}, the three
spin-gradient channels satisfy
\begin{align}
\mathcal{L}[Y]
&=-\frac{I}{3k_BT}(\bm{c}\cdot\bm{\Omega}),
\label{eq:Y_integral_eq}\\
\mathcal{L}[Z_i]
&=-\frac{I}{k_BT}(\bm{c}\times\bm{\Omega})_i,
\label{eq:Zi_integral_eq}\\
\mathcal{L}[Z_{ij}]
&=-\frac{I}{k_BT}\STF{c_j \Omega_i}.
\label{eq:Zij_integral_eq}
\end{align}
Together with Eqs.~\eqref{eq:lambda0_formula}--\eqref{eq:lambda2_formula},
these equations define $\lambda_0$, $\lambda_1$, and $\lambda_2$, hence
$(\alpha,\beta,\gamma)$ through Eq.~\eqref{eq:abg_from_lambda}.

At this point the formal closure of the spin-diffusion sector is complete in the
structural sense: the irreducible driving fields, the linear equations, and the
moment formulas are all explicit.

\subsection{Rotational-viscosity sector}
The rotational-viscosity sector is subtler. As Eq.~\eqref{eq:ci_cj_W_zero}
showed, the symmetric kinetic stress kernel $mc_ic_j$ cannot generate an axial
stress. The coefficient $\eta_r$ therefore belongs to the collisional-transfer
part of the total stress. Denote the corresponding linear stress functional by
$\mathfrak{S}_{ij}[\phi]$. In the present axial channel this gives
\begin{equation}
\sigma_{ij}^{(\intc,1)}=\mathfrak{S}_{ij}[X_ka_k].
\label{eq:Sfunctional_axial}
\end{equation}
By isotropy, parity, and linearity this map must take the form
\begin{equation}
\mathfrak{S}_{ij}[X_ka_k]=-2\eta_rA_{ij}.
\label{eq:etar_def_functional}
\end{equation}
The corresponding response function satisfies a linear equation
\begin{equation}
\mathcal{L}[X_i]=\mathcal{S}^{(\ax)}_i,
\label{eq:Xi_integral_eq}
\end{equation}
where $\mathcal{S}^{(\ax)}_i$ denotes the net axial source produced by the
collisional-transfer mechanism and by the retained-spin extended
manifold.

Equations \eqref{eq:etar_def_functional} and \eqref{eq:Xi_integral_eq} are the
formal first-order definition of $\eta_r$. To evaluate $\eta_r$ for a specific
rough-sphere collision model one still needs an explicit collisional-transfer
representation for $\mathfrak{S}_{ij}$ and the associated collision bracket.
This is precisely the point at which the antisymmetric problem becomes harder
than the ordinary symmetric stress problem.

\HLTXT_BLUE{A coefficient-complete evaluation of $\eta_r$ would require constructing the collisional-transfer part of the Cauchy stress for the finite-size rough-sphere collision operator and then evaluating the antisymmetric axial bracket associated with $X_i$. In an Irving--Kirkwood or Enskog-type formulation this stress contains contact-transfer contributions supported on the collision line segment between particle centers. Projecting the antisymmetric part of that bilinear functional onto the retained-spin axial mode is the direct bracket calculation. The present paper does not perform that calculation; instead, Sec.~\ref{subseq:dilutegasestmrotvisc} estimates $\eta_r$ by the equivalent low-density homogeneous spin-relaxation route.}

\subsection{What is closed and what is not}
It is worth isolating the logical status of the first-order calculation.
\begin{itemize}[leftmargin=2em]
\item The one-particle balance identities are fully derived, and the exact
finite-size antisymmetric-stress torque channel is made explicit.
\item The first-order constitutive \emph{form} is fully derived.
\item The spin-diffusion coefficient problem is formally closed by
Eqs.~\eqref{eq:Y_integral_eq}--\eqref{eq:Zij_integral_eq} and
\eqref{eq:lambda0_formula}--\eqref{eq:lambda2_formula}.
\item The rotational-viscosity channel is structurally closed, but its
coefficient-level evaluation still requires an explicit collisional-transfer
bracket.
\end{itemize}
This is the reason for presenting the rough-sphere formulas in the next section
as controlled dilute-gas estimates.

\section{Perfectly rough hard spheres} \label{seq:perfectlyrhs}
\subsection{Collision rule}
We now specialize to identical perfectly rough elastic hard spheres of diameter
$a$, mass $m$, and moment of inertia $I$. Introduce the standard reduced
moment-of-inertia parameter
\begin{equation}
K:=\frac{4I}{ma^2}.
\label{eq:def_K}
\end{equation}
Let the precollision states be $(\bm{v},\bm{\omega})$ and
$(\bm{v}_1,\bm{\omega}_1)$, and let $\bm{k}$ be the unit vector along the line
of centers at contact. Define the relative translational velocity
\begin{equation}
\bm{g}:=\bm{v}_1-\bm{v},
\label{eq:def_g}
\end{equation}
and the spin sum and difference
\begin{equation}
\bm{\Omega}_+:=\bm{\omega}+\bm{\omega}_1,
\qquad
\bm{\Omega}_-:=\bm{\omega}-\bm{\omega}_1.
\end{equation}
A convenient form of the collision map is
\begin{align}
\bm{v}'&=\bm{v}+\bm{M},
&
\bm{v}_1'&=\bm{v}_1-\bm{M},
\label{eq:rough_collision_v}\\
\bm{\omega}'&=\bm{\omega}-\bm{N},
&
\bm{\omega}_1'&=\bm{\omega}_1-\bm{N},
\label{eq:rough_collision_omega}
\end{align}
with
\begin{align}
\bm{M}
&=\frac{K}{K+1}\left[
\bm{g}
-\frac{a}{2}\bm{k}\times\bm{\Omega}_+
+\frac{1}{K}(\bm{k}\cdot\bm{g})\bm{k}
\right],
\label{eq:M_def}\\
\bm{N}
&=\frac{2}{aK}\bm{k}\times\bm{M}.
\label{eq:N_def}
\end{align}
These formulas encode conservation of linear momentum, total angular momentum,
and kinetic energy together with the no-slip condition at contact; see, for
example, Refs.~\cite{DahlerSather1963,CondiffLuDahler1965,McCoySandlerDahler1966}.

The relative velocity of the contact points has the same normal component as the
center-to-center relative velocity,
\begin{equation}
\bm{k}\cdot\bm{g}_c=\bm{k}\cdot\bm{g},
\label{eq:gc_normal}
\end{equation}
so the collision rate factor is $\Theta(\bm{k}\cdot\bm{g})(\bm{k}\cdot\bm{g})$.
This identity is used repeatedly below.

\HLTXT_ORANGE{The perfectly rough elastic hard-sphere model is used here as a reference model, not as a claim that most granular or suspension flows are elastic. Its role is to isolate the rotational exchange mechanism in the simplest setting where number, linear momentum, total angular momentum, and total kinetic energy are conserved in each collision. Elasticity removes homogeneous cooling and energy injection from the coefficient calculation, and perfect roughness removes an additional tangential restitution parameter. More realistic inelastic rough granular gases are essential for many applications and involve translational--rotational temperature nonequipartition, cooling, driving, and non-Gaussian translation--rotation correlations. Those effects have been studied extensively in granular kinetic theory and simulations. They are not included here because the present objective is to identify the retained-spin micropolar channels in the cleanest elastic reference problem before adding dissipative granular physics.}

\HLTXT_ORANGE{The restriction to elastic collisions distinguishes the present reference calculation from the extensive literature on inelastic rough granular gases, including kinetic theories for rough inelastic disks and spheres, studies of translation--rotation correlations, and recent treatments of rough granular transport coefficients and inertial suspensions \cite{JenkinsRichman1985,Brilliantov2007,Kranz2009,GayenAlam2008,SantosKremerGarzo2010,GomezGonzalezGarzo2020,KremerSantos2022}. Those works address dissipative physics that is deliberately excluded from the present elastic coefficient estimate.}

\subsection{Dilute-gas estimate of the rotational viscosity $\eta_r$} \label{subseq:dilutegasestmrotvisc}
\subsubsection{Homogeneous reference state}
Consider a spatially homogeneous state with zero mean flow, uniform density,
uniform temperature, and a small uniform mean spin $\bm{\omega}_{\eh}(t)$. A
natural leading-order reference distribution is the shifted Maxwellian
\begin{equation}
f_{\eh}(\bm{v},\bm{\omega};t)
=
n
\left(\frac{m}{2\pi k_BT}\right)^{3/2}
\left(\frac{I}{2\pi k_BT}\right)^{3/2}
\exp\!\left[-\frac{mv^2}{2k_BT}-\frac{I|\bm{\omega}-\bm{\omega}_{\eh}|^2}{2k_BT}\right].
\label{eq:fh_def}
\end{equation}
It satisfies
\begin{equation}
\int \bm{\omega}f_{\eh}\,\dd^3v\,\dd^3\omega=n\bm{\omega}_{\eh},
\qquad
\rho J\bm{\omega}_{\eh}=nI\bm{\omega}_{\eh}.
\label{eq:fh_moment}
\end{equation}
Our goal is to compute the linear relaxation rate of $\bm{\omega}_{\eh}$.

\subsubsection{Collisional production of the spin density}
In a binary collision the change in the pair spin sum is
\begin{equation}
\Delta\bm{\Omega}_+
:=(\bm{\omega}'+\bm{\omega}_1')-(\bm{\omega}+\bm{\omega}_1)
=-2\bm{N}.
\label{eq:DeltaOmegaplus_def}
\end{equation}
Using Eq.~\eqref{eq:N_def},
\begin{equation}
\Delta\bm{\Omega}_+
=-\frac{4}{aK}\bm{k}\times\bm{M}.
\end{equation}
Substituting Eq.~\eqref{eq:M_def} and using the vector identity
$\bm{k}\times(\bm{k}\times\bm{a})=\bm{k}(\bm{k}\cdot\bm{a})-\bm{a}$ gives
\begin{align}
\Delta\bm{\Omega}_+
&=-\frac{4}{a(K+1)}\bm{k}\times\bm{g}
+\frac{2}{K+1}\Bigl[\bm{k}(\bm{k}\cdot\bm{\Omega}_+)-\bm{\Omega}_+\Bigr].
\label{eq:DeltaOmegaplus_formula}
\end{align}
The first term does not contribute after the incoming-hemisphere average over
$\bm{k}$: for fixed $\bm{g}$, the integral
$\int_{\mathbb{S}^2}\dd\bm{k}\;\Theta(\bm{k}\cdot\bm{g})(\bm{k}\cdot\bm{g})(\bm{k}\times\bm{g})$
vanishes by axial symmetry around $\bm{g}$. The second term produces the
linear relaxation.

The dilute collisional production of the spin density may be estimated directly
from the pair-collision change in the spin sum and symmetrization over the two
collision partners. This gives
\begin{equation}
\partial_t(nI\omega_{\eh,i})
=\frac{a^2I}{2}
\int\dd^3v\,\dd^3v_1\,\dd^3\omega\,\dd^3\omega_1
\int_{\mathbb{S}^2}\dd\bm{k}\;
\Theta(\bm{k}\cdot\bm{g})(\bm{k}\cdot\bm{g})
\,f_{\eh}f_{\eh,1}\,\Delta\Omega_{+,i}.
\label{eq:etar_spin_production}
\end{equation}
Here $f_{\eh,1}=f_{\eh}(\bm{v}_1,\bm{\omega}_1;t)$.

\subsubsection{Angular average}
For fixed $\bm{g}$, symmetry implies that the hemisphere tensor integral must be
of the form
\begin{equation}
\int_{\mathbb{S}^2}\dd\bm{k}\;\Theta(\bm{k}\cdot\bm{g})(\bm{k}\cdot\bm{g})k_ik_j
=A(|\bm{g}|)\delta_{ij}+B(|\bm{g}|)\hat g_i\hat g_j,
\label{eq:tensor_ansatz}
\end{equation}
where $\hat{\bm{g}}=\bm{g}/|\bm{g}|$. The coefficients are found by taking the
trace and the projection along $\hat{\bm{g}}$.

The trace gives
\begin{align}
3A+B
&=\int_{\mathbb{S}^2}\dd\bm{k}\;\Theta(\bm{k}\cdot\bm{g})(\bm{k}\cdot\bm{g})
\nonumber\\
&=|\bm{g}|\int_0^{2\pi}\dd\phi\int_0^{\pi/2}\cos\theta\sin\theta\,\dd\theta
=\pi|\bm{g}|.
\end{align}
Projecting with $\hat g_i\hat g_j$ gives
\begin{align}
A+B
&=\int_{\mathbb{S}^2}\dd\bm{k}\;\Theta(\bm{k}\cdot\bm{g})(\bm{k}\cdot\bm{g})(\hat{\bm{g}}\cdot\bm{k})^2
\nonumber\\
&=|\bm{g}|\int_0^{2\pi}\dd\phi\int_0^{\pi/2}\cos^3\theta\sin\theta\,\dd\theta
=\frac{\pi}{2}|\bm{g}|.
\end{align}
Solving these two equations gives
\begin{equation}
A=B=\frac{\pi}{4}|\bm{g}|,
\end{equation}
so
\begin{equation}
\int_{\mathbb{S}^2}\dd\bm{k}\;\Theta(\bm{k}\cdot\bm{g})(\bm{k}\cdot\bm{g})k_ik_j
=\frac{\pi|\bm{g}|}{4}(\delta_{ij}+\hat g_i\hat g_j).
\label{eq:angular_tensor_average}
\end{equation}
Subtracting the identity tensor gives
\begin{equation}
\int_{\mathbb{S}^2}\dd\bm{k}\;\Theta(\bm{k}\cdot\bm{g})(\bm{k}\cdot\bm{g})(k_ik_j-\delta_{ij})
=\pi|\bm{g}|\left(-\frac34\delta_{ij}+\frac14\hat g_i\hat g_j\right).
\label{eq:angular_tensor_shifted}
\end{equation}
For fixed $|\bm{g}|$, an isotropic average over the direction of $\bm{g}$ uses
$\avg{\hat g_i\hat g_j}_{\hat{\bm{g}}}=\delta_{ij}/3$, giving
\begin{equation}
\avg{\int_{\mathbb{S}^2}\dd\bm{k}\;\Theta(\bm{k}\cdot\bm{g})(\bm{k}\cdot\bm{g})(k_ik_j-\delta_{ij})}_{\hat{\bm{g}}}
=-\frac{2\pi}{3}|\bm{g}|\,\delta_{ij}.
\label{eq:angular_tensor_directional}
\end{equation}
A subsequent Maxwellian average over $|\bm{g}|$ then gives
\begin{equation}
\avg{\int_{\mathbb{S}^2}\dd\bm{k}\;\Theta(\bm{k}\cdot\bm{g})(\bm{k}\cdot\bm{g})(k_ik_j-\delta_{ij})}_{0}
=-\frac{2\pi}{3}\,\avg{|\bm{g}|}_{0}\,\delta_{ij}.
\label{eq:angular_tensor_isotropic}
\end{equation}
The mean relative speed of two Maxwellian particles at temperature $T$ is
\begin{equation}
\avg{|\bm{g}|}_{0}=4\sqrt{\frac{k_BT}{\pi m}},
\label{eq:mean_rel_speed}
\end{equation}
with the derivation recorded in Appendix~\ref{app:maxwell_moments}.

\subsubsection{Relaxation rate and constitutive matching}
Using Eq.~\eqref{eq:DeltaOmegaplus_formula} in
Eq.~\eqref{eq:etar_spin_production}, discarding the $\bm{k}\times\bm{g}$ term
whose incoming-hemisphere average vanishes by axial symmetry around $\bm{g}$,
and using the fact that $\bm{g}$ and $\bm{\Omega}_+$ are statistically
independent in the reference state, we obtain
\begin{align}
\partial_t(nI\omega_{\eh,i})
&=\frac{a^2I}{2}\cdot\frac{2}{K+1}
\int f_{\eh}f_{\eh,1}
\left[
\int_{\mathbb{S}^2}\dd\bm{k}\;\Theta(\bm{k}\cdot\bm{g})(\bm{k}\cdot\bm{g})(k_ik_j-\delta_{ij})
\right]\Omega_{+,j}
\dd\Gamma\,\dd\Gamma_1
\nonumber\\
&=-\frac{2\pi a^2I}{3(K+1)}\avg{|\bm{g}|}_{0}
\int f_{\eh}f_{\eh,1}\,\Omega_{+,i}\,\dd\Gamma\,\dd\Gamma_1.
\end{align}
Since $\avg{\bm{\Omega}_+}=2\bm{\omega}_{\eh}$ and
$\int f_{\eh}f_{\eh,1}\,\dd\Gamma\,\dd\Gamma_1=n^2$, we obtain
\begin{equation}
\partial_t\bm{\omega}_{\eh}=-\nu_{\mathrm{spin}}\bm{\omega}_{\eh},
\qquad
\nu_{\mathrm{spin}}=
\frac{16\sqrt{\pi}}{3(K+1)}na^2\sqrt{\frac{k_BT}{m}}.
\label{eq:nu_spin_final}
\end{equation}

Now compare this microscopic decay law with the homogeneous zero-vorticity limit
of the macroscopic spin equation,
\begin{equation}
\rho J\partial_t\bm{\omega}_{\eh}=-4\eta_r\bm{\omega}_{\eh}.
\label{eq:macro_spin_relax}
\end{equation}
\HLTXT_BLUE{The calculation above gives a homogeneous spin-relaxation rate. To convert it into a dilute estimate of the rotational viscosity, we match this microscopic relaxation law to the homogeneous zero-vorticity limit of the macroscopic spin equation. Thus the following step is a constitutive matching estimate, not a direct evaluation of the full axial collisional-transfer bracket. Matching Eqs.~\eqref{eq:nu_spin_final} and \eqref{eq:macro_spin_relax} gives}
\begin{align}
\eta_r
&=\frac{\rho J}{4}\nu_{\mathrm{spin}}
=\frac{nI}{4}\nu_{\mathrm{spin}}
\nonumber\\
&=\frac{4\sqrt{\pi}}{3(K+1)}n^2Ia^2\sqrt{\frac{k_BT}{m}}
=\frac{\sqrt{\pi}K}{3(K+1)}n^2ma^4\sqrt{\frac{k_BT}{m}}.
\label{eq:etar_final}
\end{align}
This is positive for every $K>0$, vanishes with $K\to0$, and scales as $n^2$,
which is the characteristic signature of a collisional exchange coefficient.

\subsection{Dilute-gas estimate of the transverse spin-diffusion combination $\beta+\gamma$} \label{subsec:dilutegasestofthetran}
\subsubsection{Setup and first-order ansatz}
We next estimate the transverse combination $\beta+\gamma$. Consider a state
with zero mean flow, uniform density, uniform temperature, and a slowly varying
transverse mean spin,
\begin{equation}
\bm{u}=\bm{0},
\qquad
n=\mathrm{const.},
\qquad
T=\mathrm{const.},
\qquad
\bm{\omega}_0(y)=\omega_{0z}(y)\,\bm{e}_z.
\label{eq:betagamma_setup}
\end{equation}
Then $\Div\bm{\omega}_0=0$ and the constitutive law reduces to
\begin{equation}
m_{zy}=(\beta+\gamma)\,\partial_y\omega_{0z}.
\label{eq:mzy_constitutive}
\end{equation}
A local Maxwellian reference state is
\begin{equation}
f^{(0)}(y,\bm{c},\bm{\omega})
=
n
\left(\frac{m}{2\pi k_BT}\right)^{3/2}
\left(\frac{I}{2\pi k_BT}\right)^{3/2}
\exp\!\left[-\frac{mc^2}{2k_BT}-\frac{I|\bm{\omega}-\bm{\omega}_0(y)|^2}{2k_BT}\right].
\label{eq:f0_betagamma}
\end{equation}
Define
\begin{equation}
\Omega_i:=\omega_i-\omega_{0i}(y),
\qquad
g_s:=\partial_y\omega_{0z}.
\label{eq:gs_def}
\end{equation}
Differentiating Eq.~\eqref{eq:f0_betagamma} with respect to $y$ gives
\begin{equation}
\partial_yf^{(0)}
=\frac{I}{k_BT}\Omega_zf^{(0)}g_s,
\qquad
c_y\partial_yf^{(0)}
=\frac{I}{k_BT}c_y\Omega_zf^{(0)}g_s.
\label{eq:betagamma_source}
\end{equation}
Hence the relevant microscopic mode is
\begin{equation}
X_{zy}:=\Omega_zc_y.
\label{eq:Xzy_def}
\end{equation}
In a leading transport-relaxation or first-Sonine approximation we write
\begin{equation}
f^{(1)}_{\mathrm{sd}}=A_mX_{zy}f^{(0)}g_s.
\label{eq:f1sd_ansatz}
\end{equation}
If $\nu_m$ denotes the relaxation rate of the mode $X_{zy}$, then projection of
the linearized equation gives
\begin{equation}
A_m=-\frac{I}{k_BT\,\nu_m}.
\label{eq:Am_def}
\end{equation}
The corresponding couple stress is
\begin{align}
m_{zy}
&=-I\int \Omega_zc_yf^{(1)}_{\mathrm{sd}}\,\dd^3c\,\dd^3\omega
\nonumber\\
&=-IA_mg_s\int \Omega_z^2c_y^2f^{(0)}\,\dd^3c\,\dd^3\omega
\nonumber\\
&=-IA_mg_sn\avg{\Omega_z^2}\avg{c_y^2}
\nonumber\\
&=-IA_mg_sn\left(\frac{k_BT}{I}\right)\left(\frac{k_BT}{m}\right)
\nonumber\\
&=\frac{nIk_BT}{m\nu_m}g_s.
\label{eq:mzy_from_num}
\end{align}
Matching with Eq.~\eqref{eq:mzy_constitutive} gives
\begin{equation}
\beta+\gamma=\frac{nIk_BT}{m\nu_m}.
\label{eq:betagamma_from_num}
\end{equation}
The task is therefore reduced to computing $\nu_m$.

The relaxation calculation below concerns the single transverse component
$X_{zy}$. By isotropy, one may equivalently evaluate any fixed Cartesian
component pair of the same type, and we may take $(i,j)=(z,y)$ throughout. For
notational compactness we nevertheless keep the symbols $i$ and $j$ in the
pair-mode algebra, but from this point onward they are fixed component labels
rather than dummy indices: Einstein summation is not being used for $i$ or $j$
(that is, there is no sum over $i$ or $j$). Only explicitly repeated auxiliary
indices such as $\ell$ are summed.

\subsubsection{Pair mode and collision kinematics}
Define the pair mode for such a fixed component pair
\begin{equation}
Y_{ij}:=\Omega_ic_j+\Omega_{1i}c_{1j}.
\label{eq:Yij_pair}
\end{equation}
Introduce the sum/difference variables
\begin{equation}
\bm{V}:=\bm{c}+\bm{c}_1,
\qquad
\bm{g}:=\bm{c}_1-\bm{c},
\qquad
\bm{\Omega}_\pm:=\bm{\Omega}\pm\bm{\Omega}_1.
\label{eq:pair_variables}
\end{equation}
Then
\begin{equation}
Y_{ij}=\frac12\Omega_{+,i}V_j-\frac12\Omega_{-,i}g_j.
\label{eq:Yij_split}
\end{equation}
For the rough collision rule, $\bm{V}$ and $\bm{\Omega}_-$ are invariants, while
\begin{align}
\Delta\bm{\Omega}_+
&=-\frac{4}{a(K+1)}\bm{k}\times\bm{g}
+\frac{2}{K+1}\Bigl[\bm{k}(\bm{k}\cdot\bm{\Omega}_+)-\bm{\Omega}_+\Bigr],
\label{eq:DeltaOmegaplus_again}\\
\Delta\bm{g}
&=-\frac{2K}{K+1}\bm{g}
+\frac{aK}{K+1}\bm{k}\times\bm{\Omega}_+
-\frac{2}{K+1}(\bm{k}\cdot\bm{g})\bm{k}.
\label{eq:Deltag_formula}
\end{align}
Equation \eqref{eq:Deltag_formula} follows directly from
$\bm{g}'=\bm{v}_1'-\bm{v}'=\bm{g}-2\bm{M}$ and Eq.~\eqref{eq:M_def}.
Therefore
\begin{equation}
\Delta Y_{ij}
=\frac12(\Delta\Omega_{+,i})V_j
-\frac12\Omega_{-,i}(\Delta g_j).
\label{eq:DeltaYij}
\end{equation}

In the homogeneous Maxwellian reference state with zero mean spin, the random
vectors $\bm{V}$, $\bm{g}$, $\bm{\Omega}_+$, and $\bm{\Omega}_-$ are mutually
independent and satisfy
\begin{equation}
\avg{V_j^2}=\avg{g_j^2}=2v_T^2,
\qquad
\avg{\Omega_{+,i}^2}=\avg{\Omega_{-,i}^2}=2\omega_T^2,
\label{eq:VgOmega_variances}
\end{equation}
with
\begin{equation}
v_T^2:=\frac{k_BT}{m},
\qquad
\omega_T^2:=\frac{k_BT}{I}.
\end{equation}
Using Eq.~\eqref{eq:Yij_split},
\begin{align}
\avg{Y_{ij}^2}
&=\frac14\avg{\Omega_{+,i}^2}\avg{V_j^2}
+\frac14\avg{\Omega_{-,i}^2}\avg{g_j^2}
\nonumber\\
&=\frac14(2\omega_T^2)(2v_T^2)+\frac14(2\omega_T^2)(2v_T^2)
=2v_T^2\omega_T^2.
\label{eq:Yij2}
\end{align}

\subsubsection{Relaxation rate as a collision bracket}
The linear relaxation rate of this fixed-component mode is defined by
\begin{equation}
\nu_m
=-a^2n\,
\frac{\displaystyle
\avg{\int_{\mathbb{S}^2}\dd\bm{k}\;\Theta(\bm{k}\cdot\bm{g})(\bm{k}\cdot\bm{g})\,\Delta Y_{ij}\,Y_{ij}}_0}
{\avg{Y_{ij}^2}_0}.
\label{eq:num_def}
\end{equation}
To evaluate the numerator for this fixed component pair, introduce
\begin{equation}
A_{ij}:=\frac12\Omega_{+,i}V_j,
\qquad
B_{ij}:=-\frac12\Omega_{-,i}g_j,
\end{equation}
so that $Y_{ij}=A_{ij}+B_{ij}$. Then
\begin{equation}
\Delta Y_{ij}Y_{ij}=(\Delta A_{ij})A_{ij}+(\Delta B_{ij})B_{ij}
+(\Delta A_{ij})B_{ij}+(\Delta B_{ij})A_{ij}.
\end{equation}
The mixed terms vanish after averaging because they involve independent
zero-mean variables. Thus the numerator splits into two sectors.

\paragraph*{Sector A: the $\Omega_+V$ contribution.}
Using Eq.~\eqref{eq:DeltaYij},
\begin{equation}
(\Delta A_{ij})A_{ij}=\frac14V_j^2(\Delta\Omega_{+,i})\Omega_{+,i}.
\end{equation}
The term proportional to $\bm{k}\times\bm{g}$ in
Eq.~\eqref{eq:DeltaOmegaplus_again} does not contribute because its hemisphere
average vanishes by axial symmetry around $\bm{g}$. Therefore,
\begin{align}
\avg{\int \Theta(\bm{k}\cdot\bm{g})(\bm{k}\cdot\bm{g})(\Delta A_{ij})A_{ij}}_0
&=\frac14\avg{V_j^2}_0\frac{2}{K+1}\sum_{\ell=1}^3\avg{\Omega_{+,i}\Omega_{+,\ell}}_0
\avg{\int \Theta(\bm{k}\cdot\bm{g})(\bm{k}\cdot\bm{g})(k_ik_\ell-\delta_{i\ell})}_0
\nonumber\\
&=\frac14(2v_T^2)\frac{2}{K+1}(2\omega_T^2)\left(-\frac{2\pi}{3}\avg{|\bm{g}|}_{0}\right)
\nonumber\\
&=-\frac{16\sqrt{\pi}}{3(K+1)}v_T^3\omega_T^2.
\label{eq:sectorA_result}
\end{align}
In the second line we used the Maxwellian-averaged tensor identity
\eqref{eq:angular_tensor_isotropic} together with Eq.~\eqref{eq:mean_rel_speed}.

\paragraph*{Sector B: the $\Omega_-g$ contribution.}
Here
\begin{equation}
(\Delta B_{ij})B_{ij}=\frac14\Omega_{-,i}^2(\Delta g_j)g_j.
\end{equation}
From Eq.~\eqref{eq:Deltag_formula},
\begin{equation}
(\Delta g_j)g_j
=-\frac{2K}{K+1}g_j^2
+\frac{aK}{K+1}(\bm{k}\times\bm{\Omega}_+)_jg_j
-\frac{2}{K+1}(\bm{k}\cdot\bm{g})k_jg_j.
\end{equation}
The term involving $\bm{\Omega}_+$ averages to zero because $\avg{\bm{\Omega}_+}=0$.
Thus
\begin{equation}
(\Delta g_j)g_j
\to
-\frac{2K}{K+1}g_j^2
-\frac{2}{K+1}(\bm{k}\cdot\bm{g})k_jg_j.
\label{eq:Delta_g_gj_reduced}
\end{equation}
The needed angular integrals are
\begin{align}
\int_{\mathbb{S}^2}\dd\bm{k}\;\Theta(\bm{k}\cdot\bm{g})(\bm{k}\cdot\bm{g})
&=\pi|\bm{g}|,
\label{eq:ang_scalar}\\
\int_{\mathbb{S}^2}\dd\bm{k}\;\Theta(\bm{k}\cdot\bm{g})(\bm{k}\cdot\bm{g})^2k_j
&=\frac{\pi}{2}|\bm{g}|g_j,
\label{eq:ang_vector}
\end{align}
with the derivation recorded in Appendix~\ref{app:hemisphere_integrals}. Using
Eq.~\eqref{eq:Delta_g_gj_reduced} we obtain
\begin{align}
\int_{\mathbb{S}^2}\dd\bm{k}\;\Theta(\bm{k}\cdot\bm{g})(\bm{k}\cdot\bm{g})(\Delta g_j)g_j
&=-\frac{2K}{K+1}\pi|\bm{g}|g_j^2
-\frac{2}{K+1}\cdot\frac{\pi}{2}|\bm{g}|g_j^2
\nonumber\\
&=-\pi\frac{2K+1}{K+1}|\bm{g}|g_j^2.
\end{align}
Therefore
\begin{align}
\avg{\int \Theta(\bm{k}\cdot\bm{g})(\bm{k}\cdot\bm{g})(\Delta B_{ij})B_{ij}}_0
&=\frac14\avg{\Omega_{-,i}^2}_0\left[-\pi\frac{2K+1}{K+1}\avg{|\bm{g}|g_j^2}_0\right]
\nonumber\\
&=\frac14(2\omega_T^2)\left[-\pi\frac{2K+1}{K+1}\cdot\frac{32}{3\sqrt{\pi}}v_T^3\right]
\nonumber\\
&=-\frac{16\sqrt{\pi}}{3}\frac{2K+1}{K+1}v_T^3\omega_T^2.
\label{eq:sectorB_result}
\end{align}
Here we used the Maxwellian moment
\begin{equation}
\avg{|\bm{g}|g_j^2}_0=\frac13\avg{|\bm{g}|^3}_0=\frac{32}{3\sqrt{\pi}}v_T^3,
\label{eq:gg2_moment}
\end{equation}
proved in Appendix~\ref{app:maxwell_moments}.

\paragraph*{Total rate and coefficient.}
Adding Eqs.~\eqref{eq:sectorA_result} and \eqref{eq:sectorB_result} gives
\begin{equation}
\avg{\int \Theta(\bm{k}\cdot\bm{g})(\bm{k}\cdot\bm{g})\Delta Y_{ij}Y_{ij}}_0
=-\frac{32\sqrt{\pi}}{3}v_T^3\omega_T^2.
\label{eq:num_total}
\end{equation}
Remarkably, the explicit $K$ dependence cancels. Substituting
Eqs.~\eqref{eq:Yij2} and \eqref{eq:num_total} into Eq.~\eqref{eq:num_def} gives
\begin{equation}
\nu_m
=\frac{16\sqrt{\pi}}{3}na^2\sqrt{\frac{k_BT}{m}}.
\label{eq:num_final}
\end{equation}
Finally, Eq.~\eqref{eq:betagamma_from_num} yields
\begin{align}
\beta+\gamma
&=\frac{nIk_BT}{m\nu_m}
\nonumber\\
&=\frac{3I}{16\sqrt{\pi}a^2}\sqrt{\frac{k_BT}{m}}
=\frac{3K}{20}\eta_0a^2,
\label{eq:betagamma_final}
\end{align}
where
\begin{equation}
\eta_0:=\frac{5}{16a^2}\sqrt{\frac{mk_BT}{\pi}}
\label{eq:eta0_def}
\end{equation}
is the dilute smooth-hard-sphere shear viscosity.

\subsection{Brief note on the symmetric stress sector}
The symmetric stress sector for perfectly rough spheres is classical and may be
imported from the first-Sonine Pidduck calculation
\cite{Pidduck1922,CondiffLuDahler1965,McCoySandlerDahler1966}. In the perfectly
rough elastic limit one may write
\begin{equation}
\eta=\eta_0\frac{6(1+K)^2}{6+13K},
\qquad
\xi=\eta_0\frac{(1+K)^2}{10K}.
\label{eq:eta_xi_pidduck}
\end{equation}
These formulas are not rederived here because the algebra is standard and the
main purpose of the present paper is to spell out the less familiar
antisymmetric and spin-diffusion channels.

\section{Targeted event-driven molecular-dynamics checks}
\label{sec:edmd_checks}
\HLTXT_ORANGE{The rough-sphere estimates obtained above can be compared with targeted numerical diagnostics for the same perfectly rough elastic hard spheres. The purpose of this section is intentionally limited. The EDMD data are not intended as a parameter-free validation of the full Chapman--Enskog coefficient problem. Rather, they provide a posteriori checks of selected signatures of the retained-spin closure, primarily the homogeneous spin-relaxation mechanism used to estimate $\eta_r$ and the qualitative structure of a finite-$k$ retained-spin response. A full validation would require independent evaluations of the transport coefficients, for example by Green--Kubo formulas or by nonequilibrium protocols designed to isolate each coefficient. That separate transport-simulation program is not attempted here.}

Specifically, the expanded homogeneous dataset is used to assess the predicted collisional $n^2$
scaling and roughness dependence of $\eta_r$, while the linear finite-$k$
transverse retained-spin response associated with $\beta+\gamma$ is used as a
qualitative diagnostic. All runs in this section use event-driven molecular
dynamics in a periodic cubic box with $N=8192$ particles and $32$
statistically independent seeds per batch.

\subsection{Protocol and observables}
For the homogeneous runs we prepare states with zero mean flow and a spatially
uniform mean spin along $z$. The ensemble-averaged spin signal
$\overline{\omega}_z(t)$ is fitted by a single exponential,
\begin{equation}
\overline{\omega}_z(t)\approx \overline{\omega}_z(0)e^{-\nu_{\mathrm{spin}}^{\mathrm{EDMD}}t},
\qquad
\eta_r^{\mathrm{EDMD}}=\frac{nI}{4}\nu_{\mathrm{spin}}^{\mathrm{EDMD}},
\label{eq:edmd_homogeneous_fit}
\end{equation}
where the second relation is the macroscopic matching condition already used in
Eq.~\eqref{eq:macro_spin_relax}. For the finite-$k$ runs we initialize a single
transverse retained-spin mode and record the complex Fourier amplitudes
$\hat{\omega}_z(t)$ and $\hat{\zeta}_z(t)$. These are fitted to the linearized
two-field system
\begin{equation}
\partial_t
\begin{pmatrix}
\hat{\zeta}_z\\
\hat{\omega}_z
\end{pmatrix}
=
\begin{pmatrix}
-\dfrac{(\eta+\eta_r)k^2}{\rho} & \dfrac{2\eta_r k^2}{\rho}\\[0.8em]
\dfrac{2\eta_r}{\rho J} & -\dfrac{(\beta+\gamma)k^2+4\eta_r}{\rho J}
\end{pmatrix}
\begin{pmatrix}
\hat{\zeta}_z\\
\hat{\omega}_z
\end{pmatrix},
\label{eq:edmd_transverse_matrix}
\end{equation}
which implies the internal consistency condition $A_{12}=Jk^2A_{21}$ for the
off-diagonal entries of the fitted generator. In the updated homogeneous
dataset, we consider a baseline A1 run at $(\phi,K)=(0.020,0.400)$, a density
sweep A2 at fixed $K=0.400$ over $0.005\le \phi\le 0.050$, and a roughness
sweep A3 at fixed $\phi=0.020$ over $0.050\le K\le 1.000$. Throughout this
section the error bars shown in the figures are bootstrap $16\%$--$84\%$
intervals obtained from the seed ensemble.

\subsection{Homogeneous spin relaxation and the rotational viscosity}
\begin{figure}[t]
\centering
\includegraphics[width=0.78\textwidth]{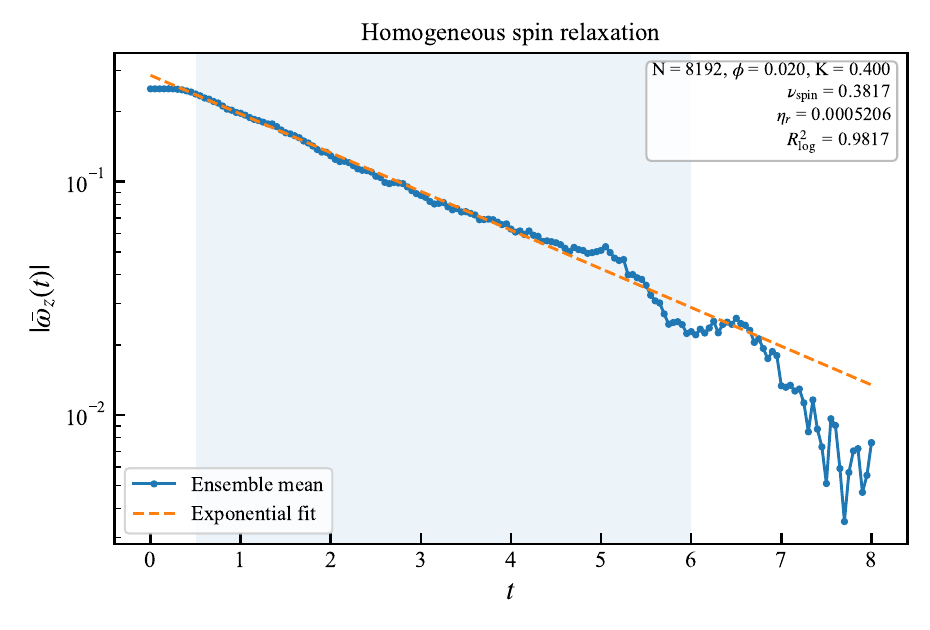}
\caption{Homogeneous spin relaxation in EDMD for perfectly rough elastic hard
spheres at $N=8192$, $\phi=0.020$, and $K=0.400$. The solid curve is the
ensemble-averaged magnitude of the mean spin, and the dashed line is the
single-exponential fit used to extract $\nu_{\mathrm{spin}}$ and hence
$\eta_r$ via Eq.~\eqref{eq:edmd_homogeneous_fit}. The shaded interval marks the
fit window.}
\label{fig:edmd_homogeneous_spin}
\end{figure}

Figure~\ref{fig:edmd_homogeneous_spin} shows the baseline homogeneous-spin run
at $\phi=0.020$ and $K=0.400$. Over the shaded fit interval the ensemble mean is
well described by a single exponential, giving
$\nu_{\mathrm{spin}}^{\mathrm{EDMD}}=0.38165$,
$\eta_r^{\mathrm{EDMD}}=5.21\times10^{-4}$, and
$R^2_{\log}=0.9817$. This is the cleanest numerical check in the paper because
it targets precisely the homogeneous relaxation mechanism from which
Eq.~\eqref{eq:etar_final} was inferred.

\begin{figure}[t]
\centering
\includegraphics[width=0.78\textwidth]{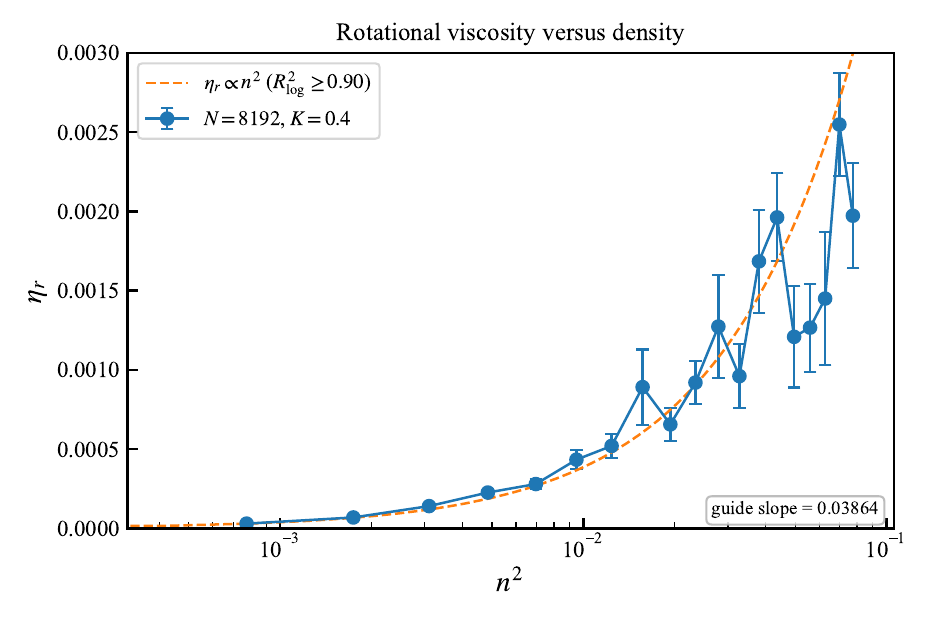}
\caption{Rotational viscosity extracted from homogeneous-spin EDMD runs as a
function of $n^2$ at fixed $K=0.400$ for the expanded density sweep
$0.005\le\phi\le0.050$. The dashed line is a guide proportional to $n^2$,
normalized through the better-conditioned points ($R^2_{\log}\ge 0.90$).}
\label{fig:edmd_eta_density}
\end{figure}

The density sweep at fixed $K=0.400$ is shown in
Fig.~\ref{fig:edmd_eta_density}. The expanded sweep now spans
$0.005\le\phi\le0.050$, i.e. a full two decades in $n^2$. Over the low- to
intermediate-density part of the set the extracted $\eta_r$ follows the
collisional $n^2$ guide closely, rising from $3.04\times10^{-5}$ at
$\phi=0.005$ to $1.27\times10^{-3}$ at $\phi=0.030$, while the single-exponential
quality remains high ($R^2_{\log}\approx0.95$ or better for most points up to
$\phi\approx0.03$). At higher densities the data still grow overall but become
visibly more scattered and less perfectly monotone; the fit quality drops to
$R^2_{\log}=0.8945$ at $\phi=0.035$, $0.7378$ at $\phi=0.0475$, and $0.5900$ at
$\phi=0.050$. We therefore interpret Fig.~\ref{fig:edmd_eta_density} as strong
support for the dilute-to-moderate-density $n^2$ trend, together with a clear
indication that the highest-density points are already feeling departures from a
simple single-exponential relaxation picture.

\begin{figure}[t]
\centering
\includegraphics[width=0.78\textwidth]{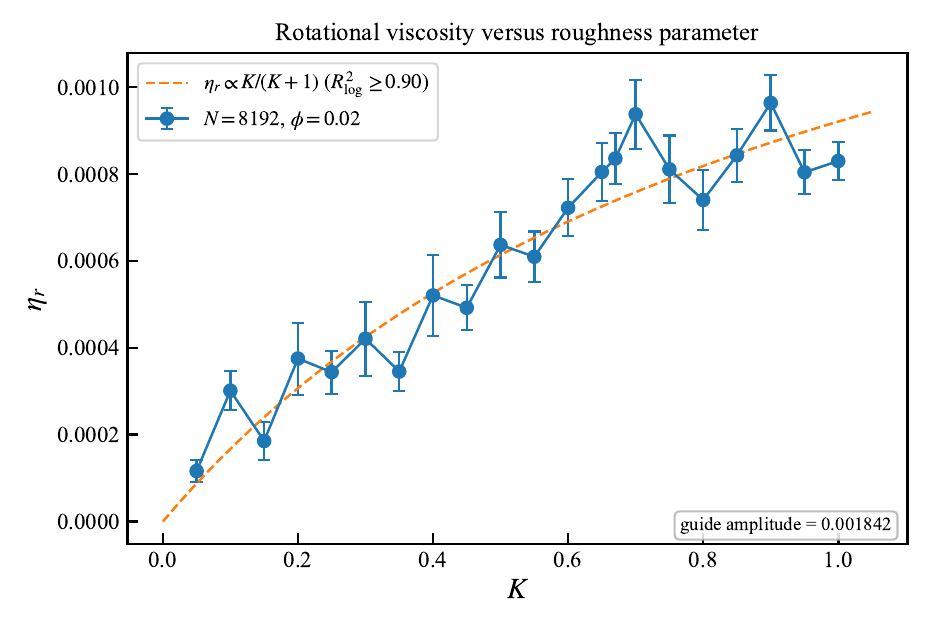}
\caption{Rotational viscosity extracted from homogeneous-spin EDMD runs as a
function of the reduced moment-of-inertia parameter $K$ at fixed
$\phi=0.020$ for the expanded roughness sweep $0.050\le K\le1.000$. The dashed
curve is a guide proportional to $K/(K+1)$, normalized through the
better-conditioned points ($R^2_{\log}\ge 0.90$).}
\label{fig:edmd_eta_K}
\end{figure}

Figure~\ref{fig:edmd_eta_K} shows the expanded roughness sweep at fixed
$\phi=0.020$ over $0.050\le K\le1.000$. The extracted $\eta_r$ grows overall
from $1.16\times10^{-4}$ at $K=0.05$ to values of order
$(0.8$--$1.0)\times10^{-3}$ for $K\approx0.65$--$1.0$. With the denser sampling,
the expected $K/(K+1)$ dependence is more clearly visible than before: aside
from the smallest-$K$ cases, the points track the dashed guide reasonably well
and approach a broad high-$K$ plateau of the expected magnitude. The least
well-conditioned runs occur at $K=0.05$ and $0.10$, where
$R^2_{\log}=0.7912$ and $0.7323$, with a milder dip $R^2_{\log}=0.8500$ at
$K=0.20$; for most points with $K\ge0.25$, however, the log-linearity is
strong. We therefore read Fig.~\ref{fig:edmd_eta_K} as a qualitatively
convincing confirmation of the roughness dependence predicted by
Eq.~\eqref{eq:etar_final}.

Taken together, Figs.~\ref{fig:edmd_homogeneous_spin}--\ref{fig:edmd_eta_K}
now provide a substantially broader homogeneous benchmark for the dilute-gas
rotational-viscosity estimate. The density sweep covers a full two-decade range
in $n^2$, and the roughness sweep resolves the interval from weak to order-one
roughness. Within the better-conditioned subset of the data, both central
signatures of Eq.~\eqref{eq:etar_final}---collisional $n^2$ scaling and growth
with roughness---are clearly visible. The deviations concentrated at the
highest densities and the smallest $K$ are better interpreted as limits of the
simple fitting ansatz than as contradictions of the underlying dilute-gas
picture.

\subsection{Finite-$k$ transverse retained-spin decay}
\begin{figure}[t]
\centering
\includegraphics[width=0.92\textwidth]{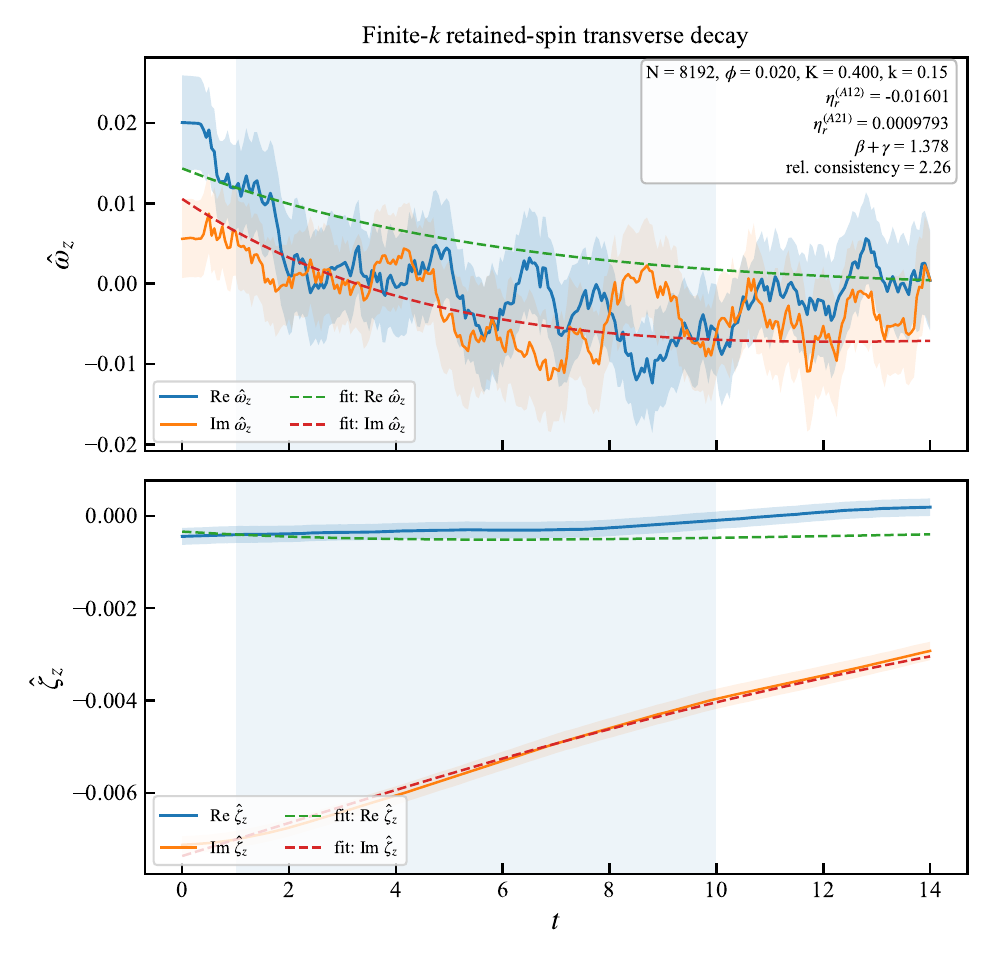}
\caption{Representative finite-$k$ retained-spin EDMD run at $N=8192$,
$\phi=0.020$, $K=0.400$, and $k=0.14996$ (batch B2). The solid curves are the
ensemble-averaged complex Fourier amplitudes and the dashed curves are the
best-fit linear two-field model from Eq.~\eqref{eq:edmd_transverse_matrix}.
The $\hat{\zeta}_z$ component is reproduced reasonably well, but the
$\hat{\omega}_z$ component remains noisy and the off-diagonal extractions of
$\eta_r$ are not mutually consistent. In the present data this figure should
therefore be interpreted as a qualitative diagnostic rather than as a
high-precision coefficient measurement.}
\label{fig:edmd_transverse_decay}
\end{figure}

The interpretation of the finite-$k$ runs is essentially unchanged by the
present update. Figure~\ref{fig:edmd_transverse_decay} shows a representative finite-$k$ run at
$\phi=0.020$, $K=0.400$, and $k=0.14996$. Two independent batches (B1 and B2)
produced the same qualitative picture, and Fig.~\ref{fig:edmd_transverse_decay}
shows the cleaner B2 realization. The $\hat{\zeta}_z$ channel is captured
reasonably well by the linear two-field fit, with $R^2_{\zeta}=0.9017$, whereas
the $\hat{\omega}_z$ channel remains noisy and is not well described by a
single fitted generator ($R^2_{\omega}=-0.0026$ in the plotted case).
Correspondingly, the two off-diagonal estimates of the rotational viscosity do
not agree:
$\eta_r^{(A_{12})}=-1.60\times10^{-2}$ and
$\eta_r^{(A_{21})}=9.79\times10^{-4}$, giving a relative cross-consistency
measure of $2.26$. The companion B1 batch yields the same qualitative outcome,
with $R^2_{\zeta}=0.9156$, $R^2_{\omega}=0.0021$, and similarly incompatible
off-diagonal $\eta_r$ estimates.

\HLTXT_ORANGE{The finite-$k$ EDMD data are therefore best viewed as a diagnostic baseline, not as a coefficient-level validation of $\beta+\gamma$. They indicate that a coupled retained-spin/vorticity relaxation channel can be observed in the simulated bulk response, but the present fitted transients do not provide a parameter-free measurement of the off-diagonal micropolar couplings. In the context of the present paper this limitation is deliberate: the homogeneous simulations check the scaling signatures of the low-density $\eta_r$ estimate, while the finite-$k$ runs identify what a future dedicated response study would have to measure independently.}

\clearpage
\section{Discussion and conclusion}
We summarize the main logical points of the derivation.

First, the one-particle balance identities follow directly from the
Boltzmann--Curtiss equation once the collision invariants are specified, while
for finite-size rough particles the exact local intrinsic-spin balance must keep
the antisymmetric-stress torque explicit. The intrinsic-spin balance is most
transparent when derived from total angular momentum and then reduced by
subtracting the orbital part.

Second, when the mean spin $\bm{\omega}_0$ is retained explicitly, the relevant
first-order Chapman--Enskog calculation is a generalized one in the sense of
extended hydrodynamics. In the bookkeeping adopted here, the residual axial
relaxation of the retained-spin manifold is placed at $O(\eps)$ within the
collisional-transfer channel, so the retained-spin reference manifold is
quasi-equilibrium rather than a strictly instantaneous elimination of spin.

Third, the irreducible decomposition of the first-order source cleanly separates
the shear, bulk, spin-diffusion, and axial mismatch channels. The symmetric
one-particle kinetic stress contains no axial part, and the spin-gradient
sectors are excluded from it by parity. As a result, the rotational viscosity
$\eta_r$ belongs to the collisional-transfer stress torque channel.

Fourth, the first-order constitutive structure recovers the standard micropolar
form,
\begin{align*}
\rho\frac{\D\bm{u}}{\D t}
&=-\grad P
+(\eta+\eta_r)\grad^2\bm{u}
+\left(\frac{\eta}{3}+\xi-\eta_r\right)\grad(\Div\bm{u})
+2\eta_r\Curl\bm{\omega}_0
+\rho\bm{F},
\\
\rho J\frac{\D\bm{\omega}_0}{\D t}
&=(\beta+\gamma)\grad^2\bm{\omega}_0
+(\alpha+\beta-\gamma)\grad(\Div\bm{\omega}_0)
+2\eta_r(\vort-2\bm{\omega}_0)
+\rho\bm{G}.
\end{align*}
The derivation displayed here makes it easy to see where each operator comes
from.

\HLTXT_ORANGE{The bulk setting used here should also be interpreted with care. For identical circular disks or symmetric spheres, spin effects in an unbounded bulk flow may be weak or absent unless spin gradients, vorticity--spin mismatch, or boundary-induced rotation are present. This is consistent with granular-flow studies in which micropolar effects are most visible near rough boundaries, inclined planes, or shear-localized regions \cite{Mitarai2002}. The homogeneous relaxation calculation in Sec.~\ref{subseq:dilutegasestmrotvisc} deliberately prepares a uniform mean spin in order to isolate the rotational-viscosity channel, while the transverse calculation in Sec.~\ref{subsec:dilutegasestofthetran} deliberately imposes a spin gradient in order to define the corresponding couple-stress coefficient. These are coefficient probes, not claims that large couple stresses are automatically generated in every bulk flow of symmetric particles.}

\HLTXT_ORANGE{This qualification is also important in relation to Babic's averaged balance equations for granular materials \cite{Babic1997}. Babic showed that, for identical circular particles, the macroscopic couple-stress contribution associated with contact interactions can vanish under the corresponding coarse-grained averaging. The present dilute spin-diffusion estimate should not be read as contradicting that result. In Sec.~\ref{subsec:couplestressatfirstord} the leading dilute $m_{ij}$ is the nonconvective flux of intrinsic angular momentum, $m_{ij}^{(k)}=-\int I\omega_i c_j f\,\dd\Gamma$, and the coefficient $\beta+\gamma$ is defined by the response to an imposed transverse spin gradient. If no spin gradient is present, this contribution vanishes. If the contact couple-stress part cancels for identical circular particles in the sense of Babic's averaging, then the observable bulk couple stress is correspondingly absent in that setting. The role of the present calculation is narrower: it identifies the formal retained-spin transport channel and estimates its dilute kinetic scale for a rough-sphere reference model.}

Finally, for perfectly rough elastic hard spheres, the dilute-gas estimates
\eqref{eq:etar_final} and \eqref{eq:betagamma_final} provide explicit physical
scales for the antisymmetric stress and the transverse couple stress. The
rotational viscosity scales as $n^2$ because it is collisional, whereas
$\beta+\gamma$ has the scale of an ordinary kinetic transport coefficient. The
expanded EDMD checks of Sec.~\ref{sec:edmd_checks} now provide a broader
homogeneous benchmark for $\eta_r$: the denser A2 density sweep supports the
predicted $n^2$ scaling over a two-decade range in $n^2$, and the A3 roughness
sweep makes the expected $K/(K+1)$ trend visible across most of the
well-conditioned dataset. At the same time, the highest-density homogeneous
points and the finite-$k$ transverse extraction remain numerically more
delicate, so the transverse channel should still be interpreted qualitatively
rather than as a precision coefficient measurement. 
Complementary response-theoretic consequences of the retained-spin closure, including 
EDMD observability and model-discrimination tests, are treated separately in a 
companion manuscript~\cite{Tsuzuki2026CompanionRespTheoConseq}.
What remains open is the complete coefficient-level evaluation of the full axial collisional transfer
bracket and of the longitudinal combination $\alpha+\beta-\gamma$ for a
concrete microscopic collision operator. That is a natural next step, but it is
logically separate from the structural first-order derivation given here.

\appendix

\section{Hemisphere integrals used in the rough-sphere calculation}
\label{app:hemisphere_integrals}
In this appendix we derive the hemisphere integrals used in the rough-sphere
estimates.

Choose coordinates so that $\hat{\bm{g}}=\bm{e}_3$ and write
\begin{equation}
\bm{k}=(\sin\theta\cos\phi,\sin\theta\sin\phi,\cos\theta),
\qquad
0\le\theta\le\frac{\pi}{2},
\qquad
0\le\phi<2\pi.
\end{equation}
Then $\bm{k}\cdot\bm{g}=|\bm{g}|\cos\theta$ on the incoming hemisphere.
Therefore
\begin{align}
\int_{\mathbb{S}^2}\dd\bm{k}\;\Theta(\bm{k}\cdot\bm{g})(\bm{k}\cdot\bm{g})
&=|\bm{g}|\int_0^{2\pi}\dd\phi\int_0^{\pi/2}\cos\theta\sin\theta\,\dd\theta
\nonumber\\
&=\pi|\bm{g}|.
\end{align}
Next,
\begin{align}
\int_{\mathbb{S}^2}\dd\bm{k}\;\Theta(\bm{k}\cdot\bm{g})(\bm{k}\cdot\bm{g})^2k_j
&=|\bm{g}|^2\int_0^{2\pi}\dd\phi\int_0^{\pi/2}\cos^2\theta\,k_j\sin\theta\,\dd\theta.
\end{align}
By azimuthal symmetry only the component along $\hat{\bm{g}}$ survives. For
$j=3$,
\begin{align}
\int_{\mathbb{S}^2}\dd\bm{k}\;\Theta(\bm{k}\cdot\bm{g})(\bm{k}\cdot\bm{g})^2k_3
&=|\bm{g}|^2\int_0^{2\pi}\dd\phi\int_0^{\pi/2}\cos^3\theta\sin\theta\,\dd\theta
\nonumber\\
&=\frac{\pi}{2}|\bm{g}|^2.
\end{align}
Since $g_j=|\bm{g}|\hat g_j$, this is equivalent to
\begin{equation}
\int_{\mathbb{S}^2}\dd\bm{k}\;\Theta(\bm{k}\cdot\bm{g})(\bm{k}\cdot\bm{g})^2k_j
=\frac{\pi}{2}|\bm{g}|g_j.
\end{equation}
This is Eq.~\eqref{eq:ang_vector}.

The tensor integral \eqref{eq:angular_tensor_average} was derived in the main
text by symmetry plus two contractions. The same result can be checked directly
in the chosen coordinates. For example, the $33$ component is
\begin{align}
\int_{\mathbb{S}^2}\dd\bm{k}\;\Theta(\bm{k}\cdot\bm{g})(\bm{k}\cdot\bm{g})k_3^2
&=|\bm{g}|\int_0^{2\pi}\dd\phi\int_0^{\pi/2}\cos^3\theta\sin\theta\,\dd\theta
\nonumber\\
&=\frac{\pi}{2}|\bm{g}|,
\end{align}
while the $11$ and $22$ components are equal and satisfy
\begin{equation}
2I_{11}+I_{33}=\pi|\bm{g}|.
\end{equation}
Thus $I_{11}=I_{22}=\pi|\bm{g}|/4$, which again gives
Eq.~\eqref{eq:angular_tensor_average}.

\section{Maxwellian moments of relative and sum variables}
\label{app:maxwell_moments}
Let $\bm{c}$ and $\bm{c}_1$ be independent centered Maxwellian velocities with
variance $k_BT/m$ in each Cartesian component. Then
\begin{equation}
\bm{g}=\bm{c}_1-\bm{c},
\qquad
\bm{V}=\bm{c}_1+\bm{c}
\end{equation}
are independent centered Gaussians with variance $2k_BT/m$ in each component.
Hence the probability density of $g:=|\bm{g}|$ is
\begin{equation}
p(g)=4\pi g^2\left(\frac{m}{4\pi k_BT}\right)^{3/2}\exp\!\left(-\frac{mg^2}{4k_BT}\right),
\qquad g\ge0.
\end{equation}
The mean relative speed is
\begin{align}
\avg{g}
&=4\pi\left(\frac{m}{4\pi k_BT}\right)^{3/2}\int_0^\infty g^3\exp\!\left(-\frac{mg^2}{4k_BT}\right)\dd g
\nonumber\\
&=4\sqrt{\frac{k_BT}{\pi m}},
\end{align}
which is Eq.~\eqref{eq:mean_rel_speed}.

Similarly,
\begin{align}
\avg{g^3}
&=4\pi\left(\frac{m}{4\pi k_BT}\right)^{3/2}\int_0^\infty g^5\exp\!\left(-\frac{mg^2}{4k_BT}\right)\dd g
\nonumber\\
&=\frac{32}{\sqrt{\pi}}\left(\frac{k_BT}{m}\right)^{3/2}
=\frac{32}{\sqrt{\pi}}v_T^3.
\label{eq:g3_result_app}
\end{align}
By isotropy,
\begin{equation}
\avg{g\,g_j^2}=\frac13\avg{g^3}=\frac{32}{3\sqrt{\pi}}v_T^3,
\end{equation}
which is Eq.~\eqref{eq:gg2_moment}.

For the spin variables, if $\bm{\Omega}$ and $\bm{\Omega}_1$ are independent
centered Gaussians with variance $k_BT/I$ in each component, then
$\bm{\Omega}_\pm=\bm{\Omega}\pm\bm{\Omega}_1$ are independent centered
Gaussians with variance $2k_BT/I$ in each component. This gives the variances
quoted in Eq.~\eqref{eq:VgOmega_variances}.

\begin{acknowledgments}
This study was supported by JSPS KAKENHI (Grant Number 22K14177) and JST PRESTO (Grant Number JPMJPR23O7).
\end{acknowledgments}

\bibliography{main}

\end{document}